\documentclass[12pt]{iopart}

\usepackage{iopams}
\expandafter\let\csname equation*\endcsname    \relax
\expandafter\let\csname endequation*\endcsname\relax
\usepackage{mathtools}
\usepackage{dsfont}
\usepackage[breaklinks=true,colorlinks,citecolor=blue,linkcolor=blue,urlcolor=blue]{hyperref}
\usepackage{cite}

\DeclarePairedDelimiter\floor{\lfloor}{\rfloor}
\def \id {\mathds{1}}
\def \abs {\text{Abs}}
\newcommand{\norm}[1]{\lVert#1\rVert}
\newcommand{\ket}[1]{|#1\rangle}
\newcommand{\bra}[1]{\langle#1|}
\def \cov {\Sigma^{(N)}}

\newtheorem{thm}{Theorem}[section]
\newtheorem{cor}{Corollary}
\newtheorem{defn}{Definition}[section]

\begin{document}
\title{Incompatibility in Quantum Parameter Estimation} 

\author{Federico Belliardo}
\ead{federico.belliardo@sns.it}
\address{NEST, Scuola Normale Superiore, I-56126 Pisa,~Italy}

\author{Vittorio Giovannetti}
\ead{vittorio.giovannetti@sns.it}
\address{NEST, Scuola Normale Superiore and Istituto Nanoscienze-CNR, I-56126 Pisa, Italy}
\begin{abstract}
In this paper we introduce a measure of genuine quantum incompatibility in the estimation task of multiple parameters, that has a geometric character and is backed by a clear operational interpretation. This measure is then applied to some simple systems in order to track the effect of a local depolarizing noise on the incompatibility of the estimation task. A semidefinite program is described and used to numerically compute the figure of merit when the analytical tools are not sufficient, among these we include an upper bound computable from the symmetric logarithmic derivatives only. Finally we discuss how to obtain compatible models for a general unitary encoding on a finite dimensional probe.

\end{abstract}


\noindent\submitto{\NJP}

\maketitle

\section{Introduction}
\label{sec:introduction}
Quantum metrology~\cite{Giovannetti2004, Giovannetti2006, Giovannetti2011, Braun2018, SCIARRINO2020} is a special branch of quantum information theory that focuses on the possibility of using quantum effects for improving the accuracy of conventional estimation procedures. Thanks to the huge variety of potential applications (which among other include the probing of delicate biological systems~\cite{Taylor2016}, optical interferometry~\cite{Caves1981, Demkowicz2015}, gravitational wave detection~\cite{Acernese2019, Tse2019}, magnetometry~\cite{Budker2007, Koschorreck2010, Wasilewski2010, Sewell2012, Troiani2018} and atomic clocks~\cite{Ludlow2015, Louchet2010, Kessler2014}), this research field is likely to play a fundamental role in the looming quantum technology revolution. As evident from the seminal works of Holevo~\cite{Holevo} and Helstrom~\cite{Helstrom1969}, this research field can be thought as a quantum counterpart of Experimental Design~\cite{Fisher,James}. Specifically the main goal of quantum metrology is to efficiently plan different types of experiments by minimizing the invested effort to overcome noisy fluctuations that originate by fabrication errors, external fields, microscopic degrees of freedom that are only statistically taken into account, and intrinsic limitations related to the formal structure of the quantum theory itself (e.g. the Heisenberg uncertainty principle). In recent years many significant results have accumulated in the domain of multi-parameter quantum metrology~\cite{Szczykulska2016, Albarelli2020}, i.e. processes where an agent tries to recover two or more attributes of a physical system (modeled by real numbers) via properly chosen measurements. The first studies that lit up the experimental interest in this subject have been done on the joint estimation of phase and phase diffusion~\cite{Vidrighin2014, Altorio2015, Roccia2018, Szczykulska2017, Crowley2014}, on quantum imaging~\cite{Parniak2018, Pezze2017, Humphreys2013, Gagatsos2016, Knott2016, Polino2019, Ciampini2016, Zhang2017}, and on magnetometry~\cite{Baumgratz2016, Apellaniz2018}. What makes the problem intriguing is that in a purely quantum setting, due to constraints ultimately related to the incompatibility of non-commuting observables~\cite{Heinosaari2016}, it could be that an efficient experiment for the determination of one specific parameter leads to poor results in the precision of the others (while this may also be true in classical mechanics, since here the phenomenon is related to the technological limits of the experimenter, there is no reason to believe it to be fundamental). Aim of the present work is to quantify the genuine quantum incompatibility associated with the estimation task of multiple parameters. The analysis is then applied to some simple systems of qubits and qutrits in order to track the effect of a local depolarizing noise on the incompatibility of the estimation task. A semidefinite program is described and used to numerically compute the figure of merit when the analytical tools are not sufficient. Finally we notice that the strategies that allow us to codify information without incompatibility in the two-qubits scenario can be generalized to the case of a general unitary encoding on a finite dimensional probe. Before proceeding with the presentation, we add here a terminology clarification: with ``quantum parameter'' estimation we denote the task of extracting a parameter encoded on a certain given fixed state of a quantum system, while if we use ``quantum metrology'' it means that we have the possibility of choosing the probe that will undergo the encoding process. In this perspective the problem of parameter estimation is hence a sub-problem of quantum metrology. In this paper we will take the probe to be fixed and therefore we will be dealing with parameter estimation. 

An outlook of the manuscript follows. In section~\ref{sec:multiParamQuantum} we introduce the setting of quantum metrology, and isolate the form of incompatibility that we will characterize later on. In section~\ref{sec:definitionFigure} the incompatibility figure of merit for quantum estimation is defined and its well-definedness is proved in section~\ref{sec:wellposed}. The geometric interpretation of the figure of merit is presented in section~\ref{sec:geometric} and in section~\ref{sec:formalDevComp} we express it in terms of the Holevo-Cram\'er-Rao bound~\cite{Holevo, Holevo1976}, proved to be achievable thanks to the quantum central limit theorem and the quantum local asymptotic normality (QLAN)~\cite{Hayashi2006, Yamagata2013, Hayashi2008, Guta2006, Guta2007, Guta2009, Yang2019}. This allows us to compute the incompatibility via the semidefinite program (SDP) reported in~\ref{app:SDPprogram}, which is derived from the one presented in~\cite{Albarelli2019}. In section~\ref{sec:boundr} an analytic upper bound for the incompatibility is presented, and in section~\ref{sec:separable} a version of the figure of merit for separable measurements is given. Section~\ref{sec:noisyInc} is dedicated to some examples with systems of qubits and qutrits subject to local depolarizing noise, here we put at work the linear program and some peculiar behavior of the incompatibility is observed. In section~\ref{sec:enforcing} we describe three strategies to build a compatible statistical model for a quantum metrological task involving $D$-dimensional probe states. The mathematical environments Definition, Theorem, and Corollary will be used to highlight the most important concepts that we introduce.

\section{Multiparameter quantum estimation}
\label{sec:multiParamQuantum}
\subsection{Setting and definitions}
\label{sec:settingAndDef}
A prototypical example of multi-parameter quantum metrology is provided by magnetometry~\cite{Budker2007, Koschorreck2010, Wasilewski2010, Sewell2012, Troiani2018, Rondin2012} where a spin particle is used as a probe for evaluating the three components of a magnetic field $\boldsymbol{B}:= \left( B_x, B_y, B_z \right)$. In the most basic scenario the evolution of the particle is given by the unitary transformation $U_{\boldsymbol{B}} := \exp \left[ \rmi \left( B_x S_x + B_y S_y + B_z S_z \right) t \right]$ where $S_i$ for $i = x, y, z$ are the components of the spin. By measuring the evolved state of the probe we can hence try infer the values of $B_x, B_y, B_z$, following the post-processing of the measurements output. What makes this procedure truly quantum in nature is that, fixing the number of experimental repetitions, due to the non-commuting nature of the generators $S_i$, any attempt to improve the estimation accuracy of one of the cartesian components of $\boldsymbol{B}$ will have a negative impact on the accuracies of the other two~\cite{Ragy2016}. An exact formalization of this problem can be obtained by considering a more general model where one is asked to determine $d$ parameters ${\btheta} := \left( \theta_1, \theta_2, \dots, \theta_d \right) \in \Theta$ (an open subset of $\mathbb{R}^d$) that have been encoded in the input state $\rho$ of a probing quantum system via a mapping of the form 
\begin{eqnarray} \label{MAPPING} 
	\rho \rightarrow \rho_{{\btheta}} := \mathcal{E}_{{\btheta}}\left( \rho \right)\;,
\end{eqnarray}
where now $\mathcal{E}_{{\btheta}}$ is a completely positive, trace-preserving (CPT) transformation~\cite{Nielsen2010-1} which parametrically depends on~${{\btheta}}$ and which, at variance with the simplified scenario detailed at the beginning of the section, might include a noise disturbing the process. Given $N$ copies of $\rho_{{\btheta}}$ we can now try to recover the needed information by performing on them some (possibly joint) positive operator valued measure (POVM) ${\text M}_N :=\{ E_{{\boldsymbol{\hat{\theta}}}}^{(N)} \}_{{\boldsymbol{\hat{\theta}}}}$ whose elements are labelled by a classical outcome variable ${\boldsymbol{\hat{\theta}}}$ that, without loss of generality~\cite{NOTA}, can be assumed to belong to the same set $\Theta$ of ${\btheta}$. Accordingly ${\text M}_N$ can hence be thought as a operation which, starting from $\rho_{{\btheta}}^{\otimes N}$, induces a measure on $\Theta$, defined by the conditional probability distribution
\begin{eqnarray}
	P_{{\text M}_N}({\boldsymbol{\hat{\theta}}}|{\btheta}) := \mbox{Tr} \, [ E_{{\boldsymbol{\hat{\theta}}}}^{(N)} \rho_{{\btheta}}^{\otimes N} ] \;, \label{CONDPROB} 
\end{eqnarray}
with the stochastic outcome ${\boldsymbol{\hat{\theta}}}:=({\hat{\theta}}_1, {\hat{\theta}}_2, \dots, {\hat{\theta}}_d)\in \Theta$ playing the role of the estimator of ${\btheta}$. The two most important properties of the estimator ${\boldsymbol{\hat{\theta}}}$ are the bias vector $\boldsymbol{b}^{(N)}\left({\btheta}\right) := \left(b_1 \left({\btheta}\right), b_2 \left({\btheta}\right), \cdots b_d \left({\btheta}\right)\right)$, of components
\begin{eqnarray} \label{newdefb} 
	b_i \left({\btheta}\right) := \mathbb{E} \, [ {\hat{\theta}}_i ]- \theta_i\;,
\end{eqnarray}
and the mean square error (MSE) $d\times d$ matrix ${\cov}\left({\btheta}\right)$, of elements
\begin{eqnarray}
	\cov_{ij}\left({\btheta}\right) := \mathbb{E} \, [ (\hat{\theta_i} - \theta_i ) ({{\hat{\theta}}}_j - \theta_j ) ]\;,
	\label{newdefsigma}
\end{eqnarray}
with $\mathbb{E}$ representing the statistical average computed with the probability measure in~\eref{CONDPROB}. Ideally we would like to deal with estimators that are unbiased, meaning that $\boldsymbol{b}^{(N)} ({\btheta}) = 0$ for all ${\btheta} \in \Theta$, but this may not always be possible. Accordingly in what follows we shall focus on sensing, i.e. we shall measure small variations of the parameters ${\btheta}$ around a known value and assume that we are allowed to employ locally unbiased POVMs at such special point, that is measurements which bias vector $\boldsymbol{b}^{(N)} (\btheta)$ satisfy the following conditions
\begin{eqnarray} 
	\left\{ \begin{array}{ll} 
	\boldsymbol{b}^{(N)}({\btheta})\Big|_{{\btheta}={\btheta}_0} = 0 \;, & \\ \\
	\frac{\partial}{\partial \theta_j} \boldsymbol{b}^{(N)}({\btheta})\Big|_{{\btheta}={\btheta}_0}= 0\;, & \forall j=1,\cdots, d \;. 
	\end{array}
	\right. \label{LOCAMB} 
\end{eqnarray}
For these measurements the quantum Cram\'er-Rao (QCR) bound~\cite{Giovannetti2011, Paris2009} gives a limit on the precision of the sensing task, formulated as a lower bound on the associated MSE matrix,~i.e.
\begin{eqnarray}
	\cov({\btheta}) \geq \frac{F^{-1}({\btheta})}{N} \; .
	\label{eq:multiQR}
\end{eqnarray}
In this expression $F({\btheta})$ is the so called quantum Fisher information (QFI) matrix which no longer depends on the selected POVM $\text{M}_N$ and whose elements can be computed as 
\begin{eqnarray}
	F_{ij} ({\btheta}):= \frac{1}{2} \Tr \Big[ \rho_{{\btheta}} \left( L_i ({\btheta})L_j ({\btheta})+ L_j ({\btheta})L_i({\btheta}) \right) \Big] \; ,
	\label{eq:multiFish}
\end{eqnarray}
with $L_i({\btheta})$ the symmetric logarithmic derivative (SLD)~\cite{Paris2009} associated to the $i$th component of the parameter vector~${\btheta}$, i.e. the operator (possibly dependent on ${\btheta}$) fulfilling the identity
\begin{eqnarray}
	\frac{\partial \rho_{{\btheta}}}{\partial \theta_i} = \frac{1}{2} \Big( \rho_{{\btheta}} L_i({\btheta}) + L_i({\btheta}) \rho_{{\btheta}} \Big) \; .
	\label{eq:defL}
\end{eqnarray}
For a pure state $\rho_{\btheta} = \ket{\psi_{\btheta}} \! \bra{\psi_{\btheta}}$ the above equation admits as solution 
\begin{eqnarray}
	L_{i} (\btheta) = 2 \frac{\partial \rho_{{\btheta}}}{\partial \theta_i} \; ,
	\label{eq:defLpure}
\end{eqnarray}
while in general a solution is~\cite{Paris2009}
\begin{eqnarray}
	L_i (\btheta) = 2 \int_{0}^{+\infty} \rme^{- s \rho_{\btheta}} \frac{\partial \rho_{{\btheta}}}{\partial \theta_i} \rme^{- s \rho_{\btheta}} \rmd s \; .
	\label{eq:defLsolution} 
\end{eqnarray}
Throughout the paper we will assume the QFI to be limited (i.e. $\norm{{F}({\btheta})} < \infty$) and non-singular (i.e. $F({\btheta}) > 0$). In particular the last requirement imposes that the maximum value of $d$ (the number of parameters) we can allow in our study is upper bounded by $D^2 -1$ with $D$ being the dimension of the Hilbert space associated with the probing system (indeed values of $d$ greater than such limit will necessarily force a linear dependence between the SLD operators $L_i({\btheta})$, leading to a singular QFI matrix).

\subsection{Achievability of the multi-parameter QCR bound} 
\label{sec:achievability}
In general the multiparameter QCR bound~\eref{eq:multiQR} cannot be saturated, meaning that there is no locally unbiased POVM $\text{M}_N$ with a $\cov({\btheta})$ matrix equal to $F^{-1}({\btheta})/N$ or, equivalently, which is capable of saturating the inequality
\begin{eqnarray}
	\Tr \, [ G \cdot \cov({\btheta}) ] \ge \frac{1}{N} \Tr \, [ G \cdot F^{-1}({\btheta}) ] := \frac{C_{\text{S}} \left(G ,{\btheta} \right)}{N}\; ,
	\label{eq:multiCRG}
\end{eqnarray}
for all choices of a positive weight matrix $G\ge0$. This is the form of metrological incompatibility that will be extensively studied in this paper. In order to better appreciate the meaning of this, suppose that we are interested in the estimation of an analytic function $f \in \mathcal{C}^\omega \left( \Theta \right)$ of the unknown parameters vector ${\btheta}$. The function $f$ will be evaluated on the estimator ${\boldsymbol{\hat{\theta}}}$ extracted from the observations. By expanding to first order the expectation value of $f ({\boldsymbol{\hat{\theta}}} ) - f \left( {\btheta} \right)$ we get the expression for the error
\begin{eqnarray}
	\varepsilon &:=& \mathbb{E} \, [ ( f ( {\boldsymbol{\hat{\theta}}} ) - f({{\btheta}}) ) ^ 2 ] \\ &\simeq&
	\sum_{i,j} \mathbb{E} \, [ \partial_i f (\btheta) ( {{\hat{\theta}}}_i - \theta_i ) \partial_j f (\btheta) ( {{\hat{\theta}}}_j - \theta_j ) ] \nonumber \\ &=& \sum_{i,j} \partial_i f (\btheta) \partial_j f (\btheta)
	\mathbb{E} \, [ ( {{\hat{\theta}}}_i - \theta_i ) ( {{\hat{\theta}}}_j - \theta_j ) ] \; ,
\end{eqnarray}
which can be equivalently written as:
\begin{eqnarray}
	\varepsilon = \langle \partial f (\btheta) | \cov (\btheta) | \partial f (\btheta) \rangle = \Tr \, [ G (\btheta) \cdot \cov(\btheta) ] \; ,
\end{eqnarray}
where we introduced the rank-$1$ weight matrix $G_{ij} (\btheta) = \partial_i f (\btheta) \partial_j f (\btheta) = \ket{\partial f (\btheta)} \! \bra{\partial f (\btheta)}$, with $\ket{\partial f (\btheta)} \in \mathbb{R}^3$. Written in this form we can now use~\eref{eq:multiCRG} to cast a bound on the accuracy of the estimation of $f \left( {\boldsymbol{{\theta}}}\right)$. As a matter of fact a rank-$1$ $G$ can always be thought as the weight matrix of some function $f({{\btheta}})$. We will see that according to our definitions a rank-$1$ $G$ manifests no incompatibility, indeed we will see that the error associated to a single tangent vector $\ket{\partial f (\btheta)}$ on the statistical manifold can saturate the ultimate QFI (this can be understood e.g. from the upper bound~\eref{eq:usefulBound} discussed in section~\ref{sec:boundr} below, which, for $G$ rank-$1$, collapses to $C_{\text{S}} \left(G ,{{\btheta}}\right)$). On the contrary the gap manifests itself when the weight matrix $G$ is at least rank-$2$. This situations arises as we try to estimate at the same time multiple functions of the parameters ${{\btheta}}$, named $f_1 ( {{\btheta}} )$, $f_2 ( {{\btheta}} )$ , \dots, $f_K ( {{\btheta}} )$, which could also just be the components $\theta_1, \theta_2, \dots, \theta_d$ of the vector ${{\btheta}}$. To each of the functions we associate a weight $g_i \ge 0$, then the total error is the weighted sum of the errors for the estimation of each $f_i \left( {{\btheta}} \right)$, i.e.
\begin{eqnarray}
	\varepsilon := \sum_{i = 1}^{K} g_i \Tr \, [ \ket{\partial f_i (\btheta)} \! \bra{\partial f_i (\btheta)} \cov ] = \Tr \, [ G (\btheta) \cdot \cov ], \;
\end{eqnarray}
with $G (\btheta) := \sum_{i = 1}^{K} g_i \ket{\partial f_i (\btheta)} \! \bra{\partial f_i (\btheta)} \ge 0$.

\section{Incompatibility measure}
\label{sec:figOfMerit}
In this section we introduce a figure of merit to gauge the incompatibility of multi-parameter estimation procedures,
which is based on the assumption that the agent is allowed to perform on the probes arbitrary locally unbiased POVMs. After showing its well-definedness we clarify its interpretation in the framework of information geometry. We then provide a linear program to compute this incompatibility measure and an analytical upper bound. The figure of merit is then generalized to separable measurements.

\subsection{Definition of the figure of merit.}
\label{sec:definitionFigure}
Given the encoding~\eref{MAPPING} and a generic weight matrix $G$, from~\eref{eq:multiCRG} it follows that a bona-fide evaluation of the precision attainable with a locally unbiased POVM $\text{M}_N$ can be obtained by considering the ratio
\begin{eqnarray}
	r_N \left( G, \text{M}_N , {\btheta} \right)
	 := \frac{N \Tr \left[ G \cdot \cov({\btheta}) \right]}{\Tr \left[ G \cdot F^{-1}({\btheta}) \right]} \ge 1,
	\label{eq:preTrivial}
\end{eqnarray}
where ${\cov}({\btheta})$ is the MSE matrix~\eref{newdefsigma} associated with 
$\text{M}_N$. As indicated by the notation the quantity~(\ref{eq:preTrivial}) exhibits an explicitly functional dependence on $G$ and $\text{M}_N$ which we remove by considering the term
\begin{defn}(Incompatibility figure of merit for $N$ probes)
	\begin{eqnarray}
	r_N ({\btheta}):= \inf_{\text{M}_N\in\mathcal{M}_N^{\text{(LU)}}} \sup_{G \ge 0} 	r_N \left( G, \text{M}_N,
	{\btheta}\right) \; ,
	\label{eq:definitionR}
	\end{eqnarray}
\end{defn}
where now $\mathcal{M}_N^{\text{(LU)}}$ indicates the set of locally unbiased POVM on $N$ copies of the probes. For any given elements $\text{M}_N$ of $\mathcal{M}_N^{\text{(LU)}}$ the $\sup_{G \ge 0}$ selects the weight matrix that has the reachable precision $\Tr \left[ G \cdot \cov({\btheta}) \right]$ as far away from the information content $\Tr \left[ G \cdot F^{-1}({\btheta}) \right]/N$ as possible. Then we minimize on $\text{M}_N \in \mathcal{M}_N^{\text{(LU)}}$ to compute the best worst case scenario, as in a typical min-max definition~\cite{NOTADEP}. The figure of merit $r_N ({\btheta})$ quantifies the competition between optimal measurements for different parameters, and has a clear operational meaning. Because of the QCR bound in~\eref{eq:preTrivial} we have $r_N({\btheta}) \ge 1$ and the $N$ define a fully compatible model only when $r_N({\btheta}) = 1$. This is true if and only if $\exists \, \text{M}_N\in\mathcal{M}_N^{\text{(LU)}}$ (possibly dependent on ${{\btheta}}$) for which in~\eref{eq:preTrivial} equality holds $\forall \, G$. On the contrary $r_N({\btheta}) > 1$ indicates the presence of incompatibility and happens if and only if $\forall \, \text{M}_N\in \mathcal{M}_N^{\text{(LU)}}$ $\exists \, G\geq 0$ such that in~\eref{eq:preTrivial} the strict inequality holds. In the asymptotic scenario of infinitely many probes available we introduce
\begin{defn}(Incompatibility figure of merit)
	\begin{eqnarray}
	\underline{r}({\btheta}) := \liminf_{N \rightarrow \infty} r_N({\btheta}) \; ,
	\label{eq:defr}
	\end{eqnarray}
\end{defn}
which always exists and from $r_N({\btheta})$ inherits the property $\underline{r}({\btheta}) \ge 1$. In particular in this case we have $\underline{r}({\btheta}) = 1$ if and only if there exists a sequence of $\text{M}_N\in \mathcal{M}_N^{\text{(LU)}}$, which, for all $G \ge 0$, allows us to saturate the inequality~\eref{eq:preTrivial} asymptotically in $N$. It is worth noticing that the incompatibility figure of merit could be defined for locally asymptotic covariant measurements (LAC)~\cite{Hayashi2006, Yang2019} as well, and it would be exactly equal to $\underline{r} (\btheta)$, see~\ref{app:LACfig} for the details.

\subsection{Well-definedness of the figure of merit}
\label{sec:wellposed}
We now briefly show that $r_N (\btheta)$ in~\eref{eq:definitionR} is invariant under reparametrization and therefore a well-defined quantity. This translates to $\underline{r} (\btheta)$, which is therefore a well defined property of the statistical manifold (see section~\ref{sec:geometric}). Consider a reparametrization ${{\btheta}} = {{\btheta}} \left({\bfeta} \right)$ having an invertible Jacobian $J_{ij} := \frac{\partial \theta_i ({\bfeta})}{\partial \eta_j}$. Then the MSE matrix for the parameters $\btheta$, defined in~\eref{newdefsigma}, can be written $\cov (\btheta) = J \cov (\bfeta) J^t$, where $\cov_{ij}\left(\bfeta\right) := \mathbb{E} \, [ (\hat{\eta_i} - \eta_i ) ({{\hat{\eta}}}_j - \eta_j ) ]$ is the MSE matrix for the parameters $\bfeta$. Similarly we write the inverse of the QFI matrix as $F^{-1} (\btheta) = J F^{-1} (\bfeta) J^{t}$, with $F(\bfeta)$ computed from the symmetric logarithmic derivatives $L_i(\bfeta)$, which differ from the definition in~\eref{eq:defL} in the derivatives, which are taken with respect to $\eta_i$. Its now easy to show that $r_N (\btheta) = r_N (\bfeta)$. The action of the Jacobian matrix on the MSE matrix and on the QFI can be moved on $G$, that becomes $J^t G J$ both at numerator and at denominator of the ratio in~\eref{eq:preTrivial}, while multiplying respectively $\cov (\bfeta)$ and $F^{-1}(\bfeta)$. Then we observe that the set of positive matrices is invariant under congruence for an invertible matrix, i.e. $J^t \lbrace G \ge 0 \rbrace J = \lbrace G \ge 0 \rbrace$, and therefore we get
\begin{eqnarray}
	r_N (\btheta) = \inf_{\text{M}_N\in\mathcal{M}_N^{\text{(LU)}}} \sup_{G \ge 0} \frac{N \Tr \left[ G \cdot \cov(\bfeta) \right]}{\Tr \left[ G \cdot F^{-1}(\bfeta) \right]} := r_N (\bfeta) \; .
	\label{eq:definitionReta}
\end{eqnarray}
It worth stressing that, by construction the quantity $\underline{r}({\btheta})$ only depends on the input probe state $\rho$, the encoding $\mathcal{E}_{{\btheta}}$, and the specific point of interest ${\btheta}$. It is an intrinsic property of the statistical manifold defined by the trajectories~\eref{MAPPING}. The need for a reparametrization invariant measure of incompatibility was already pointed out in~\cite{Razavian2020}, in which the figure of merit $\underline{r}(\btheta)$ was independently discovered.

\subsection{Geometric interpretation}
\label{sec:geometric}
\begin{figure}[!t]
	\centering
	\includegraphics[scale=0.6]{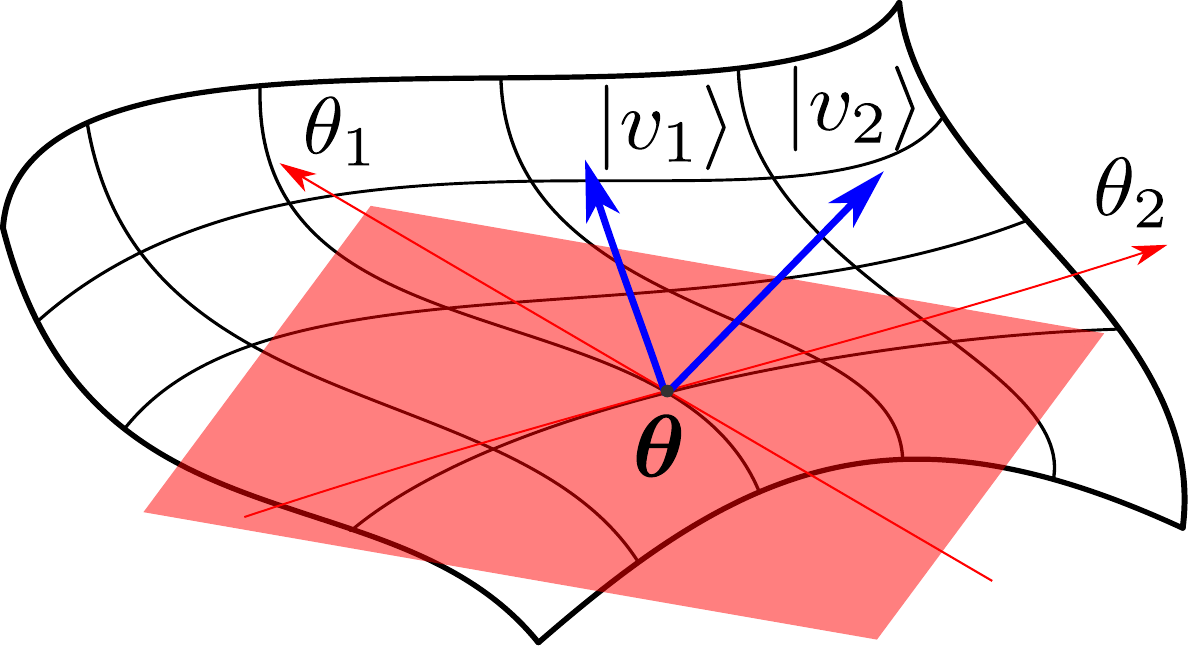}
	\caption{Representation of a 2D statistical manifold with its tangent space at a point $\btheta$ and two directions $\ket{v_1}$ and $\ket{v_2}$ on this plane.}
	\label{fig:manifold}
\end{figure}

The parameters ${{\btheta}} \in \Theta$ can be interpreted as coordinates defining via the map~(\ref{MAPPING}) a submanifold of the space of states $\mathfrak{S} \left( \mathcal{H} \right)$, called the statistical manifold. The QFI matrix, being a positive semidefinite matrix can be thought as a Riemannian metric on this manifold. This metric is generally non trivial as it explicitly depends on the coordinates ${{\btheta}}$ and may have intrinsic curvature. The QFI is said to be a distinguishability metric \cite{Braunstein1994, Braunstein1996}: given two very near states $\rho_{{\btheta}}$ and $\rho_{{{\btheta}} + d{{\btheta}}}$, their infinitesimal distance in the QFI metric is
\begin{eqnarray}
	d s ^2 := \frac{1}{4} F_{ij}({\btheta}) \rmd \theta_i \rmd \theta_j = 2 \left( 1 - \sqrt{\mathcal{F} \left( \rho_{{\btheta}}, \rho_{{{\btheta}}+d{{\btheta}}} \right)} \right),
\end{eqnarray}
which is negatively correlated with the fidelity $\mathcal{F} \left( \rho_{{\btheta}}, \rho_{{{\btheta}}+d{{\btheta}}} \right)$ between $\rho_{{\btheta}}$ and $\rho_{{{\btheta}}+d{{\btheta}}}$, defined as $\mathcal{F} \left( \rho, \sigma \right) := \left[ \Tr \left( \sqrt{\sqrt{\rho} \sigma \sqrt{\rho}} \right)\right]^2$~\cite{Nielsen2010-2}. In order to gain information about ${{\btheta}}$ it is thus better to choose the probe state $\rho$ such that in the statistical manifold the codified state $\rho_{{\btheta}}$ is highly distinguishable from its neighbors $\rho_{{{\btheta}}+d{{\btheta}}}$, and has therefore the highest statistical distance from them as possible. This picture clarifies why the inverse of the distinguishability metric, i.e. $F^{-1}({\btheta})$, gives the precision to which a single point ${\btheta}$ can be identified in $\Theta$, given the quantum state $\rho_{{\btheta}}$. For $\rho_{{\btheta}}^{\otimes N}$ the relevant metric is $F^{-1}({\btheta})/N$. When a measurement is performed and an estimator ${\boldsymbol{\hat{\theta}}}$ is chosen there is a new Riemannian metric insisting on the statistical manifold: the positive semidefinite $\cov (\btheta)$ matrix. The key question is if one can find a POVM ${\text{M}}_N\in \mathcal{M}_N^{{\text{(LU)}}}$ with a MSE metric that fully adapts to the underling quantum metric $F^{-1}({\btheta})/N$ of the manifold, i.e. if the inequality~\eref{eq:multiQR} can be saturated (at a certain point ${\btheta}$). In general this is not possible. Let us introduce a representation of $G$ as a sum of projectors $\ket{v_i} \! \bra{v_i}$, each weighted with $g_i \ge 0$, i.e. $G := \sum_{i} g_i \ket{v_i} \! \bra{v_i}$, where $\ket{v_i}$ are vectors in the tangent plane of the statistical manifold at point $\btheta$, then
\begin{eqnarray}
	\frac{1}{N} \Tr \, [ G \cdot F^{-1}({\btheta}) ] = \frac{1}{N} \sum_{i} g_i \bra{v_i} F^{-1}({\btheta}) \ket{v_i} \; .
\end{eqnarray}
According to the above expression, the information content is a weighted combination of the distinguishability of the manifold in different directions defined by $\ket{v_i}$, see figure~\ref{fig:manifold} for a 2D representation. This has to be compared with the experimental weighted distinguishability, i.e.
\begin{eqnarray}
	\Tr \, [ G \cdot \cov({\btheta}) ] = \sum_{i} g_i \bra{v_i} \cov({\btheta}) \ket{v_i}, 
\end{eqnarray}
given by a particular measurement.
The whole point of the non commutative nature of the manifold is the impossibility to saturate the distinguishability in more than one direction at the same time. By taking
\begin{eqnarray} 
	\sup_{G \ge 0} \tfrac{N \Tr \left[ G \cdot \cov({\btheta}) \right]}{\Tr \left[ G \cdot F^{-1}({\btheta}) \right] } = \sup_{g_i \ge 0, \ket{v_i}} \tfrac{N \sum_{i} g_i \bra{v_i} \cov({\btheta}) \ket{v_i}} {\sum_{i} g_i \bra{v_i} F^{-1}({\btheta}) \ket{v_i}} \; ,
\end{eqnarray}
we measure the worst case fitting of the $\cov({\btheta})$ matrix on the metric $F^{-1}({\btheta})/N$ at a point ${\btheta}$, spanning all possible sets of tangent vectors and weights. Then we minimize on the classical metric (and hence on the POVM) to find the most adapt one. By taking the asymptotic limit of infinitely many probes (through the $\liminf$) we have completed the analysis of the definition~\eref{eq:defr} from the geometrical point of view. We sum up everything and say that $\underline{r}({\btheta})$ measures, in the asymptotic scenario, the failure of finding a metric on the statistical manifold, stemming from a measurement, which fully adapts to the underlying quantum metric (in all directions) at a specific point ${\btheta}$.

\subsection{Computation of the figure of merit}
\label{sec:formalDevComp}
We would like to apply the existing results in local estimation theory to compute the incompatibility figure of merit. This requires the exchange of the $\sup$ and the $\inf$ in~\eqref{eq:definitionR}. In~\ref{app:alternativeDefinition} we do that and show
\begin{eqnarray}
	\underline{r}({\btheta}) &=& \sup_{G \ge 0} \frac{ C (G,{\btheta})}{C_{\text{S}} \left(G,{\btheta} \right)} \; ,
	\label{eq:CMI}
\end{eqnarray}
with $C_{\text{S}} \left(G,{\btheta} \right)$ defined in~(\ref{eq:multiCRG}), and with 
the numerator given by the quantity 
\begin{eqnarray}
	C (G,{\btheta}) := \lim_{n \rightarrow \infty} \min_{N \le n} \inf_{\text{M}_N\in \mathcal{M}_N^{\text{(LU)}}} N \Tr \, [ G \cdot \cov({\btheta}) ] \; . \label{DEFCFTHETA} 
\end{eqnarray}
In~\cite{Hayashi2006} it has been proved that $C(G, \btheta) = C_{\text{H}} (G, \btheta)$, where $C_{\text{H}} (G, \btheta)$ is the Holevo-Cram\'er-Rao bound functional $C_{\text{H}} (G,{\btheta})$~\cite{Hayashi8, Holevo1976}. Exploiting this facts we can use~(\ref{eq:CMI}) to deduce the following equality 
\begin{eqnarray}
	 	\underline{r}({\btheta}) = \sup_{G \ge 0} \frac{ C_{\text{H}} (G,{\btheta})}{C_\text{S} \left(G,{\btheta} \right)} \; .
	\label{eq:hcrbound}
\end{eqnarray}
In~\cite{Tsang2020} the upper bound $C_\text{H} (G,{\btheta}) \le 2 C_\text{S} (G,{\btheta})$ is given, which implies $\underline{r}({\btheta})\le 2$. Because of this it makes sense to introduce 
\begin{defn}(Incompatibility measure)
	\begin{eqnarray}
		\mathcal{I}({\btheta}) := \underline{r}({\btheta})-1 \; ,
		\label{eq:defI}
	\end{eqnarray}
\end{defn}
as a proper quantifier of incompatibility: by construction it belongs to the interval $\left[0, 1 \right]$ with $\mathcal{I}({\btheta}) = 0$ indicating full compatibility, while $\mathcal{I}({\btheta}) = 1$ maximal incompatibility. The Holevo-Cram\'er-Rao bound can be computed via the semidefinite linear program in~\cite{Albarelli2019}, which can be adapted to compute $\mathcal{I} (\btheta)$, as reported in~\ref{app:SDPprogram}.

\subsection{Upper bound on $\underline{r}(\btheta)$}
\label{sec:boundr}
In this section we propose an upper bound on $\underline{r}(\btheta)$ that relies only on the computation of the symmetric logarithmic derivatives defined in~\eref{eq:defL}. It is essentially based on $C_{\text{Z}} \left(G,{\btheta}\right)$~\cite{Suzuki2016}, a well know upper bound on $C_{\text{H}} (G,{\btheta})$, which reads
\begin{eqnarray}
	C_{\text{H}}(G,{\btheta}) &\le& C_{\text{Z}}(G,{\btheta})
	\label{eq:usefulBound} \\ \nonumber &:=& \Tr \left[ G F^{-1}({\btheta}) \right] + \Tr \abs \left[ G F^{-1}({\btheta}) 
	A({\btheta}) F^{-1}({\btheta}) \right] \; ,
\end{eqnarray}
where $A({\btheta})$ contains the expectation values of the commutators of the SLDs:
\begin{eqnarray}
	A_{ij}({\btheta}) := \frac{1}{2 \rmi} \Tr \left[ \rho_{{\btheta}} \left[ L_i({\btheta}), L_j ({\btheta})\right] \right].
\label{eq:defAnti}
\end{eqnarray}
In writing~\eref{eq:usefulBound} we used $\Tr \abs \left[ G \cdot R \right] := \Tr | \sqrt{G} R \sqrt{G} |$, with $| X | := \sqrt{X X^{\dagger}}$. 
Combining~\eref{eq:usefulBound} and~\eref{eq:hcrbound} we get
\begin{eqnarray}
	\underline{r}({\btheta}) \le 1 + \sup_{G \ge 0} \frac{\Tr \normalfont{\abs} \left[ G \cdot F^{-1}({\btheta}) A ({\btheta})F^{-1} ({\btheta})\right]}{\Tr \left[ G \cdot F^{-1}({\btheta}) \right]} := \underline{r}^\star({\btheta})\; .
	\label{eq:usefulBound2}
\end{eqnarray}
The above inequality shows that a sufficient condition to have compatibility is $A({\btheta}) = 0$. In \ref{app:proofrstar} we compute explicitly $\sup_{G \ge 0}$ in~\eref{eq:usefulBound2} and obtain
\begin{eqnarray}
	\underline{r}^{\star}({\btheta}) = 1 + \norm{F^{-\frac{1}{2}}({\btheta}) A ({\btheta})F^{-\frac{1}{2}}({\btheta})} \; ,
	\label{eq:explicitRstar}
\end{eqnarray}	
where $\norm{\cdot}$ is the operator norm. This translates to an upper bound on $\mathcal{I}({\btheta})$, i.e.
\begin{thm} (Upper bound on the incompatibility measure)
	\begin{eqnarray}
		\mathcal{I}({\btheta}) \le \mathcal{I}^\star({\btheta}) := \underline{r}^\star({\btheta}) - 1 = \norm{F^{-\frac{1}{2}}({\btheta}) A({\btheta}) F^{-\frac{1}{2}}({\btheta})}\; .
		\label{eq:Istar}
	\end{eqnarray}	
\end{thm}
This strengthen the interpretation of $A({\btheta})$ as a measure of incompatibility~\cite{Ragy2016}. The upper bound $\mathcal{I}^\star({\btheta})$ was first defined in~\cite{Carollo2019} and called $\mathcal{R}$. It has already been used as a measure of incompatibility and ``quantumness'' and applied to qubits~\cite{Razavian2020} and many-body systems~\cite{Carollo2019, Carollo2020}. By defining $\mathcal{I} (\btheta)$ we offer a more informative definition of incompatibility. It is noteworthy that for a $D$-invariant model~\cite{Suzuki2019} this bound is saturated and $\mathcal{I} (\btheta) = \mathcal{I}^\star (\btheta)$.

\subsection{Incompatibility for separable measurements}
\label{sec:separable}
We now go back to the first definition of a figure of merit presented in~\eref{eq:defr}, but consider the minimization in~\eref{eq:definitionR} to be performed only on the locally unbiased separable measurements subset $\mathcal{M}_{N}^{\text{\tiny{(LU-S)}}}$ of $\mathcal{M}_{N}^{\text{(LU)}}$ which operate locally on $\rho_{{\btheta}}^{\otimes N}$. This brings to the definitions
\begin{eqnarray}
	r_N^\text{s}({\btheta}) := \inf_{{\text{M}}_N\in \mathcal{M}_{N}^{\text{\tiny{(LU-S)}}} } \sup_{G \ge 0} 	r_N \left( G, {\text{M}}_N , {\btheta}\right) \; ,
	\label{eq:definitionRsep}
\end{eqnarray}
and
\begin{defn} (Incompatibility figure of merit for separable measurements)
	\begin{eqnarray}
		\underline{r}^\text{s}({\btheta}) := \liminf_{N \rightarrow \infty} r_N^{\text{s}}({\btheta}) \; .
		\label{eq:defrs}
	\end{eqnarray}
\end{defn}
Now we apply the result of~\cite{Gill2000}, which gives us a lower bound on the precision of the estimation with $N$ probes when we use a measurement ${\text{M}}_N \in \mathcal{M}_{N}^{\text{\tiny{(LU-S)}}}$. The bound reads
\begin{eqnarray}
	N \Tr \, [G \cdot {\cov}({\btheta}) ] \ge \frac{1}{D-1} \left( \Tr \sqrt{F^{-\frac{1}{2}}({\btheta}) G F^{-\frac{1}{2}} ({\btheta})}\right)^2 \; ,
\end{eqnarray}
where ${\cov}({\btheta})$ is the MSE matrix of ${\text{M}}_N$ and $D$ is the size of the Hilbert space of the single probe $\rho_{{\btheta}}$. This translates to a lower bound on $ r_N^\text{s} \left( G, {\text{M}}_N , {\btheta}\right)$ $\forall \, N$, i.e.
\begin{eqnarray}
	r_N^\text{s} \left( G, {\text{M}}_N ,{\btheta}\right) \ge \frac{\left( \Tr \sqrt{F^{-\frac{1}{2}}({\btheta}) G F^{-\frac{1}{2}}({\btheta})} \right)^2}{(D-1) \Tr [F^{-\frac{1}{2}}({\btheta}) G F^{-\frac{1}{2}}({\btheta}) ]} \; ,
\end{eqnarray}
which propagates to the definition of $\underline{r}^\text{s}({\btheta})$, giving
\begin{thm} (Lower bound for separable measurements)
	\begin{eqnarray}
		\underline{r}^\text{s} ({\btheta}) \ge \frac{d}{D-1} \;,
		\label{eq:GMbound}
	\end{eqnarray}
\end{thm}
where we have compute explicitly $\sup_{G \ge 0}$ using the AM-QM inequality and its saturation. Observe that the inequality~(\ref{eq:GMbound}) bares no reference to the details of the encoding process~(\ref{MAPPING}) and that it is non trivial only if the number $d$ of parameters we have to estimate is larger than or equal to $D-1$. The manipulations of \ref{app:alternativeDefinition} are valid also for the class of measurements $\mathcal{M}_{N}^{\text{\tiny{(LU-S)}}}$, because only the local unbiasedness is required in their proof. Therefore we can write
\begin{eqnarray}
	\underline{r}^\text{s}({\btheta}) = \sup_{G \ge 0} \frac{C^\text{s} \left(G,{\btheta} \right)}{C_{\text{S}} \left(G,{\btheta} \right)} \; ,
\end{eqnarray}
where now
\begin{eqnarray}
	C^{\text{s}} (G,{\btheta}) := \lim_{n \rightarrow \infty} \min_{N \le n} \inf_{ 
	{\text{M}}_N \in \mathcal{M}_{N}^{\text{\tiny{(LU-S)}}}} N \Tr \, [ G \cdot \cov ({\btheta}) ] \; .
\end{eqnarray}
At least for the case of a qubit probe ($D = 2$) the above expression allows us to exactly compute $\underline{r}^\text{s}({\btheta})$. Indeed as shown in~\cite{Hayashi8, Gill2000} for this model one has 
\begin{eqnarray}
	C^{\text{s}} (G,{\btheta}) = \left( \Tr \sqrt{F^{-\frac{1}{2}}({\btheta}) G F^{-\frac{1}{2}} ({\btheta})} \right)^2 \; ,
\end{eqnarray}
leading to
\begin{cor}
	\begin{eqnarray}
	D=2 \quad \Longrightarrow \quad \underline{r}^\text{s}({\btheta}) = d \;, 
	\end{eqnarray}
\end{cor}
which shows that in the case of a single qubit, multi-parameter estimation always exhibit incompatibility for separable locally unbiased measurements (remember that our analysis is explicitly restricted to the cases where $d\leq D^2-1=3$).

\subsection{Hierarchy of incompatibility measures}
\label{sec:hierarchy}
Whether a certain estimation process is compatible or not depends on the set of measurements $\mathcal{M}_N$ that we are allowed to perform. Consider a hierarchy of POVM sets
\begin{eqnarray}\label{GERARCHIA} 
	\mathcal{M}^{(1)}_N \subseteq \mathcal{M}^{(2)}_N \subseteq \cdots \subseteq \mathcal{M}^{(k)}_N \quad \quad \forall \, N \; ,
\end{eqnarray}
we define the figure of merit $r^{\left(i\right)}_N({\btheta})$ as in~\eref{eq:definitionR}, but taking $\mathcal{M}^{(i)}_N$ as the domain of the infimum. By construction the $r^{\left(i\right)}_N({\btheta})$ satisfy the following hierarchy of inequalities 
\begin{eqnarray}
	r^{(1)}_N({\btheta}) \ge r^{(2)}_N({\btheta}) \ge \cdots \ge r^{(k)}_N({\btheta}) \quad \quad \forall \, N \; ,
\end{eqnarray}
which carries over to
\begin{eqnarray}
	\underline{r}^{(1)} ({\btheta})\ge \underline{r}^{(2)} ({\btheta})\ge \cdots \ge\underline{r}^{(k)}({\btheta}) \; ,
	\label{eq:longHierarchy}
\end{eqnarray}
when taking the proper $N\rightarrow \infty$ limits~(\ref{eq:defr}). For example the space of separable locally unbiased measurement is a subset of the set of all locally unbiased measurements, i.e.
\begin{eqnarray}
	\mathcal{M}_{N}^{\text{\tiny{(LU-S)}}} \subseteq \mathcal{M}_N^{\text{(LU)}} \; ,
\end{eqnarray}
which means $\underline{r}^\text{s}({\btheta}) \ge \underline{r}({\btheta})$.

\section{Incompatibility of a noisy estimation task}
\label{sec:noisyInc}
In this section, by using the previously defined figures of merit $\underline{r}({\btheta}) $ in~\eqref{eq:hcrbound} and $\underline{r}^\text{s}({\btheta}) $ in~\eref{eq:defrs}, we study the incompatibility of the estimation process in a few simple cases concerning the sensing of two phases $\theta_1$ and $\theta_2$ encoded by the unitary transformation
\begin{eqnarray}
	U_{{\btheta}} := \exp \left[ \rmi \left(\theta_1 \sigma_y + \theta_2 \sigma_z \right) \right] \; ,
	\label{eq:encoding}
\end{eqnarray}
acting on individual qubits. The probes will be states of one and three qubits subject to local depolarizing noise, which is given by the map
\begin{eqnarray}
	\Lambda_\lambda (\rho) := \lambda \rho + \left(1-\lambda \right) \frac{\id}{2} \; ,
	\label{eq:depolarizing}
\end{eqnarray}
with $\lambda\in [-1/3,1]$ being a characteristic parameter of the model~\cite{KING2003}. The transformation $\Lambda_\lambda$ induces a shrinking of the qubit Bloch vector by a factor given by the modulus $|\lambda|$ which can be used to gauge the intensity of the noise. In particular for $\lambda = 1$ the map \eqref{eq:depolarizing} corresponds to the noiseless evolution, and for $\lambda = 0$ to the complete depolarization process, while negative values of $\lambda$ indicate the presence of an inversion of the Bloch sphere with respect to the origin~\cite{Rosati2018}. We are interested in investigating if the noise can force the system to a more classical behavior and therefore ensure compatibility in the estimation scenario, as it does for measurements~\cite{Heinosaari2016}. We then turn to $D$-dimensional system, and with the opportune generalizations of~\eqref{eq:encoding} and~\eqref{eq:depolarizing} we explore the upper bound $\mathcal{I}^\star (\btheta)$ in~\eqref{eq:defI} for a generic system and the incompatibility $\mathcal{I} (\btheta)$ in~\eqref{eq:Istar} for a qutrit. Notice that the chosen noise is covariant and therefore in all our examples it could be applied before or after the encoding without changing the final output $\rho_{{\btheta}}$. Table~\ref{table:table1} contains a recap of the improvements and observed phenomena in the following examples.
\begin{table}[h!]
	\centering
	\begin{tabular}{|p{1.5cm}||p{6.5cm}|p{6.5cm}|} 
		\hline
		System & Known results  & Improvements/Observed phenomena \\
		\hline \hline
		1 qubit & Computation of $\mathcal{I} (\btheta)$ $(\mathcal{R})$ for qubit tomography and two phase estimation with pure states~\cite{Razavian2020}. & Computation of $\mathcal{I} (\btheta, \lambda)$ for two phase estimation with depolarizing noise. Incompatibility with separable measurements.\\
		\hline 
		3 qubit & Computation of $\norm{A (\btheta)}_F$ and $1-C_{\text{H}} (\id)/C_\text{S} (\id)$~\cite{Albarelli2019}. & Efficient computation of $\mathcal{I} (\btheta, \lambda)$. Observation of gap between $\mathcal{I}^\star (\btheta, \lambda)$ and $\mathcal{I} (\btheta, \lambda)$ and its behavior.\\
		\hline
		1 $D$-dim & \_ & Asymmetry around $\lambda = 0$ of the upper bound $\mathcal{I}^\star (\btheta, \lambda)$.\\
		\hline
		1 qutrit & \_ & Asymmetry around $\lambda = 0$ of the incompatibility measure $\mathcal{I} (\btheta, \lambda)$.\\
		\hline
	\end{tabular}
	\caption{Recap of the examples of section~\ref{sec:noisyInc} with the improvements we propose and/or the observed relevant phenomena.}
	\label{table:table1}
\end{table}

\subsection{Incompatibility for a one-qubit probe}
\label{sec:oneQubit}
First of all we analyze the case of a single qubit probe. The fact that the figure of merit is parameterization invariant allows for an elegant exact solution of the qubit model for whatever probe state and encoded phases under depolarization noise. In this example the measure $\mathcal{I} (\btheta)$ and its upper bound $\mathcal{I}^\star (\btheta)$ will coincide. After the encoding by $U_{{\btheta}}$ in~\eref{eq:encoding}, the probe undergoes the action of the noise map $\Lambda_\lambda$ in~\eqref{eq:depolarizing}, so that its final state $\rho_{{\btheta}}$ is described by the mapping~(\ref{MAPPING}) with $\mathcal{E}_{{\btheta}}$ given by
\begin{eqnarray}
	\mathcal{E}_{{\btheta}}(\rho) := \Lambda_\lambda ( U_{{\btheta}} 
	\rho U_{{\btheta}}^\dagger ) = U_{{\btheta}} \Lambda_\lambda (
	\rho) U_{{\btheta}}^\dagger \;.
	\label{eq:statisticalModel}
\end{eqnarray}
The purity of the encoded state $\rho_{{\btheta}} = \mathcal{E}_{\btheta} \left( \rho \right)$ is independent on $\btheta$, this makes the statistical model D-invariant~\cite{Suzuki2016, Suzuki2019}, and allows us to conclude that the Holevo-Cram\'er-Rao bound $C_{\text{H}} (G,{\btheta})$ coincides with $C_{\text{Z}} (G,{\btheta})$ defined in~\eref{eq:usefulBound}, therefore the inequality \eqref{eq:usefulBound2} is saturated ($\mathcal{I}({\btheta}) = \mathcal{I}^\star({\btheta})$), and the incompatibility can be computed from the symmetric logarithmic derivatives only. We consider an arbitrary qubit probe state $\rho := \frac{1}{2} \left( \id + \boldsymbol{a} \cdot \boldsymbol{\sigma} \right)$. Its Bloch vector is $\boldsymbol{a} := \left(a_x, a_y, a_z \right)$, with $\Tr \rho^2 = \frac{1}{2} (1 + \norm{\boldsymbol{a}}^2)$. After the encoding the Bloch vector of $\rho_{\btheta} = \mathcal{E}_{\btheta} \left( \rho \right)$ is $\boldsymbol{a}_{\btheta} := \lambda \left( a_x (\btheta), a_y(\btheta), a_z (\btheta) \right)$. We can perform an implicitly defined change of variables $(\theta_1, \theta_2) \rightarrow (\alpha, \beta)$, that brings us to $\boldsymbol{a}_{(\alpha, \beta)} = \lambda \, \sqrt{2 \Tr \rho^2 -1} (\cos \alpha \cos \beta, \cos \alpha \sin \beta, \sin \alpha)$. For this model~\cite{NOTAsingular} we compute the matrices $F(\alpha, \beta)$ and the $A(\alpha, \beta)$, which are 
\begin{eqnarray*}
	F(\alpha, \beta) &=&  \sqrt{2 \Tr \rho^2 -1} \begin{pmatrix}
	\lambda^2 & 0 \\
	0 & \lambda^2 \cos^2 \alpha
	\end{pmatrix} \; , \\ A(\alpha, \beta) &=&  \sqrt{2 \Tr \rho^2 -1} \begin{pmatrix}
	0 & -\lambda^3 \cos \alpha \\
	\lambda^3 \cos \alpha & 0 
	\end{pmatrix} \; ,
\end{eqnarray*}
that substituted in~\eref{eq:Istar} give
\begin{thm} (Incompatibility measure for a depolarized qubit two phase model)
	\begin{eqnarray}
		\mathcal{I} (\btheta, \lambda) = \mathcal{I} (\alpha, \beta, \lambda) = \sqrt{2 \Tr \rho^2 - 1} \, |\lambda| \quad \forall \, \btheta\,, \; \forall \, \rho\; .
		\label{eq:Iqubitstate}
	\end{eqnarray}
\end{thm}
Equation~\eref{eq:Iqubitstate} reveals that the noise level intensity controls directly the compatibility. Indeed for fixed input the value of $\mathcal{I} (\btheta, \lambda)$ reaches its maximum in the noiseless scenario ($\lambda \rightarrow 1$) providing full incompatibility $\mathcal{I} (\btheta, \lambda) \rightarrow 1$ for pure input states. On the contrary as the noise sends $\rho_{{\btheta}}$ to the completely mixed state ($\lambda \rightarrow 0$) the codified information is dissipated and the compatibility increases, indeed $\mathcal{I} (\btheta, \lambda) \rightarrow 0$. Fundamentally the same result was discover in~\cite{Razavian2020} for qubit tomography. We finally remind the reader that, as anticipated at the end of section~\ref{sec:separable}, for a single qubit we get $\underline{r}^\text{s}({\btheta}) = d = 2$ independently on the noise. Again this result is valid $\forall \, \btheta$ and for every input probe $\rho$.

\subsection{Incompatibility for three entangled qubits}
\label{sec:threeQubit}
\begin{figure}[!t]
	\centering
	\includegraphics[scale=0.7]{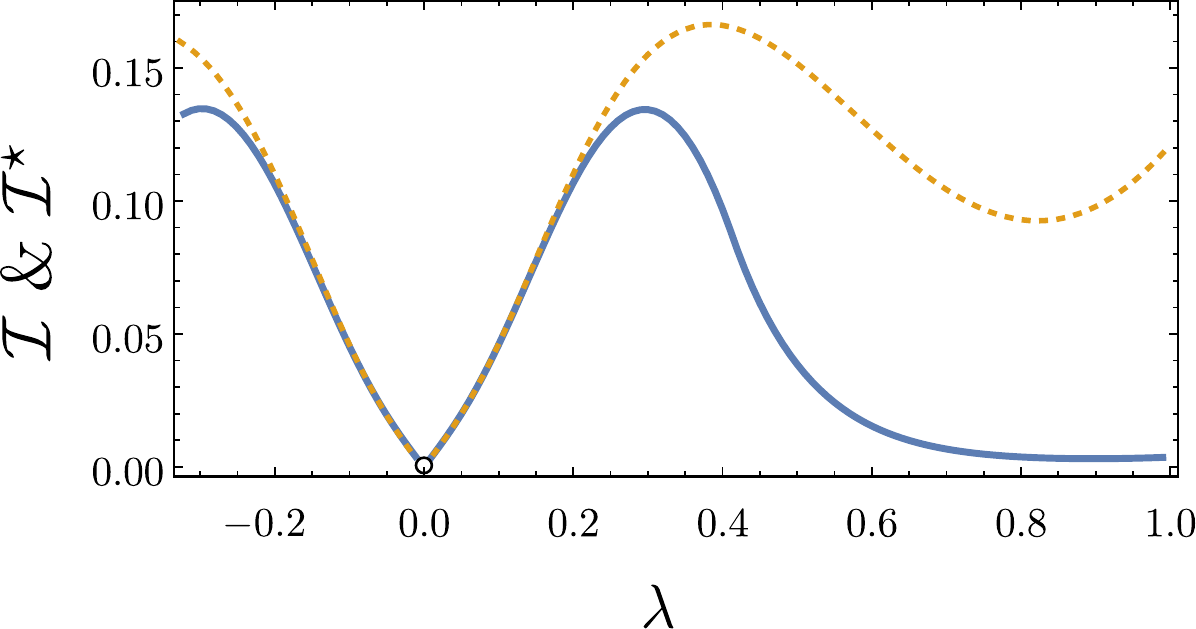}
	\caption{The orange dashed curve is the upper bound $\mathcal{I}^\star (\btheta = 0)$, defined in~\eref{eq:Istar}, computed for the encoded three qubits state~\eref{eq:threeState} as a function of the local noise intensity $\lambda$. The blue solid curve is the figure of merit $\mathcal{I} (\btheta = 0)$ defined in~\eref{eq:defI} referred to the same scenario and computed numerically as explained in \ref{app:SDPprogram}. The curves are symmetric around $\lambda = 0$. The empty point in $\lambda = 0$ indicates that at this point the information quantities are not defined.}
	\label{fig:3qubitExt}
\end{figure}

Consider now the scenario in which we have at disposal multiple copies of three entangled qubits and we codify them through $U_{{\btheta}} \otimes U_{{\btheta}} \otimes U_{{\btheta}}$, with $U_{{\btheta}}$ given in~\eref{eq:encoding}. This more complicate scenario gives us the opportunity to compute $\mathcal{I} (\btheta, \lambda)$ with the SDP and show the presence of a gap between $\mathcal{I} (\btheta, \lambda)$ and $\mathcal{I}^\star  (\btheta, \lambda)$. In this example we won't be able to compute $\underline{r}(\btheta)$ for every probe state, therefore we will concentrate on
\begin{eqnarray}
	\ket{\psi} := \frac{\ket{\psi_z} + \ket{\psi_y}}{\sqrt{2}} \; ,
	\label{eq:threeState}
\end{eqnarray}
with
\begin{eqnarray*} 
	\ket{\psi_z} &:=& \frac{1}{\sqrt{2}} \left( \ket{000} + \ket{111} \right)\;, \\
	\ket{\psi_y} &:=& \frac{1}{\sqrt{2}}\left( \ket{\phi^+ \phi^+ \phi^+} + \ket{\phi^- \phi^- \phi^-} \right),
\end{eqnarray*}
where $\ket{\phi^+}$ and $\ket{\phi^{-}}$ are the eigenvectors of $\sigma_y$ corresponding to the positive and negative eigenvalue respectively. In~\cite{Baumgratz2016} it is proved that the analogous state for the estimation of three phases with $N$ entangled qubits reaches Heisenberg scaling in the QFI in all the three parameters. At difference with the previous example, here we are able to compute the figure of merit for the probe $\ket{\psi}$ only at the point $\btheta = 0$ through numerical evaluations via the semidefinite program reported in \ref{app:SDPprogram}, these indicate a non-null $\mathcal{I} (\btheta = 0)$. We add a local depolarization noise $\Lambda_\lambda$ on each qubit and compute $\mathcal{I}({\btheta} = 0, \lambda)$ and its upper bound $\mathcal{I}^\star({\btheta} = 0, \lambda)$ as functions of $\lambda$ to see if the noise increases compatibility, the results are reported in figure~\ref{fig:3qubitExt}. $\mathcal{I}({\btheta} = 0, \lambda)$ and $\mathcal{I}^\star({\btheta} = 0, \lambda)$ have been computed for $100$ values of $\lambda$ uniformly distributed in $\left(-1/3, 1 \right)$. The addition of noise does not necessarily diminish the incompatibility, on the contrary $\mathcal{I} ({\btheta} = 0, \lambda)$ and $\mathcal{I}^\star({\btheta} = 0, \lambda)$ both display a non-monotonic behavior with respect to $|\lambda|$. This behavior of the incompatibility has already been observed in~\cite{Albarelli2019}. We notice that as the noise destroys the information codified in $\rho_{\btheta}$ both the compatibility and its upper bound $\mathcal{I}^\star (\btheta, \lambda)$ go to $0$, but this doesn't seem to be a universal behavior~\cite{Razavian2020}. We confirm a separation between $\mathcal{I}^\star (\btheta, \lambda)$ and $\mathcal{I} (\btheta, \lambda)$, that has been evidenced in~\cite{Razavian2020}, and we conjecture that $\mathcal{I} (\btheta, \lambda)-\mathcal{I}^\star (\btheta, \lambda)$ shrinks to zero as the amount of encoded information diminish, as it happens in this example for $\lambda \rightarrow 0$. In this model also the relative gap $(\mathcal{I} (\btheta, \lambda)-\mathcal{I}^\star (\btheta, \lambda))/\mathcal{I} (\btheta, \lambda)$ shrinks to zero as $\lambda \rightarrow 0$. For a generic noise this phenomenon depends on the behavior of $A(\btheta, \lambda)$ as the disturbance is increased. The figure of merit $\mathcal{I} (\btheta, \lambda)$ appears to be not correlated with the information quantities $F^{-1}_{11} (\btheta, \lambda)$ and $F^{-1}_{22} (\btheta, \lambda)$ or with the purity of the encoded state, as these measures are all monotonic in the noise $\lambda$. Also because of this we think of $\mathcal{I} (\btheta)$ as a genuine non trivial new property of the estimation process. Notice that for $\lambda = 0$, the state is unable to codify information ($F (\btheta) = 0$).

\subsection{Estimation on $D$-dimensional probes}
\label{sec:ddim}
In this section we study the incompatibility for a generic unitary encoding of $d$ parameters on a $D$-dimensional probe in $\mathfrak{S}(\mathcal{H})$, i.e.
\begin{eqnarray}
	U_{\btheta} := \exp \left( \rmi \sum_{j=1}^{d} \theta_j H_j \right) \; ,
	\label{eq:encodingDim}
\end{eqnarray}
where $H_j$ are null-trace hermitian operators acting on $\mathcal{H}$. For the estimation around $\btheta = 0$, these operators are the infinitesimal generators of the encoding. However for a generic point $\btheta \neq 0$ this is not necessarily true. As explained in \ref{app:effective}, for a given probe state, the sensing procedure around a point $\btheta \neq 0$ can however be described in terms of an effective set of new generators $H_{j}^{\text{eff}}(\btheta)$. Accordingly, since the results of the preset section are valid for estimations around $\btheta = 0$ for all possible choices of $H_j$, we can conclude that they hold true also $\forall \, \btheta$ encoded by~\eqref{eq:encodingDim}. Finally as for the noise model we replace~\eref{eq:depolarizing} with
\begin{eqnarray} 
	\Lambda_\lambda (\rho) := \lambda \rho + \left(1-\lambda \right) \frac{\id}{D}\;, 
\end{eqnarray}
which for $\lambda \in [-1/(D^2-1), 1]$ is a proper generalization of the depolarization channel for a $D$-dimensional system~\cite{KING2003,Rosati2018}.

\subsection{Incompatibility for a $D$-dimensional probe}
\label{sec:singleDdim}
\begin{figure}[!t]
	\centering
	\includegraphics[scale=0.8]{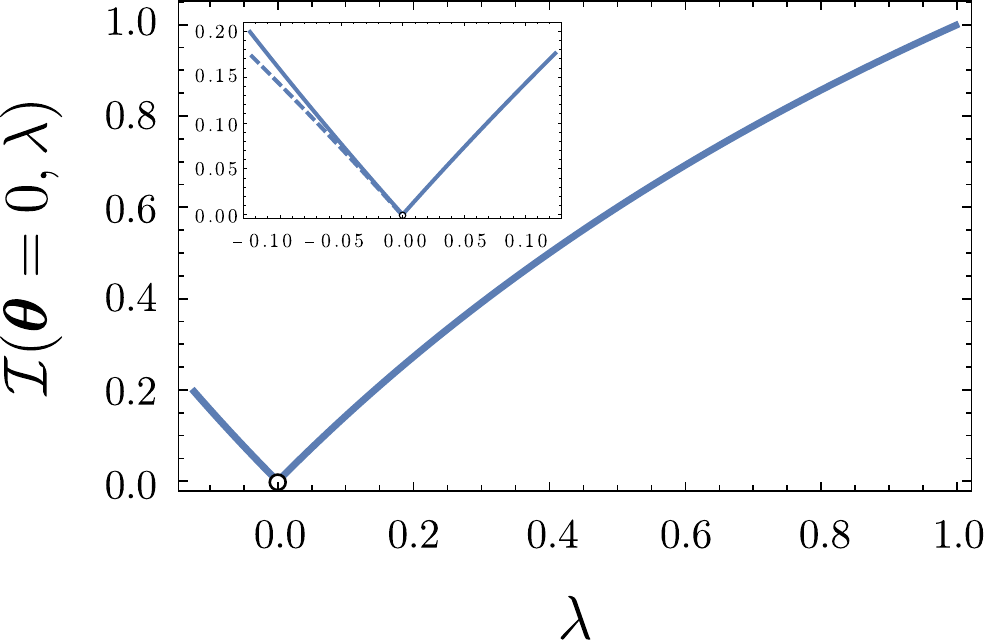}
	\caption{This curve is the incompatibility figure of merit $\mathcal{I}$ defined in~\eref{eq:defI}, for the asymptotic covariant measurements, numerically computed for the qutrit example of section~\ref{sec:singleDdim} for $\lambda \in \left( -1/8, 1 \right)$. The empty circle at $\lambda = 0$ indicates that at this point the information quantities are not defined. The dashed curve in the small insert is the mirrored figure of merit for $\lambda > 0$.}
	\label{fig:asymmetry}
\end{figure}

Let us consider a single-probe scenario where the state of the system is described by the density matrix
\begin{eqnarray}
	\rho_{{\btheta}} &:=& \Lambda_\lambda ( U_{{\btheta}} |\psi\rangle\langle\psi| U_{{\btheta}}^\dagger ) = U_{{\btheta}} \Lambda_\lambda (|\psi\rangle\langle\psi|) U_{{\btheta}}^\dagger \; \nonumber \\
	&=& \lambda |\psi_{{\btheta}}\rangle \langle\psi_{{\btheta}}|+ (1-\lambda) \frac{\id}{D}\;, 
	\label{eq:statisticalModel1}
\end{eqnarray}
with $|\psi\rangle$ being the pure input state of the system, and with $|\psi_{{\btheta}}\rangle := U_{{\btheta}}|\psi\rangle$. If we now call $L_i (\btheta)$ the symmetric logarithmic derivative associated to the parameter $\theta_i$ in the absence of noise, i.e. the SLD of $|\psi_{{\btheta}}\rangle$, given in~\eref{eq:defLpure}, then it can be seen that for $\lambda \neq 1$
\begin{eqnarray}
	L_i (\btheta, \lambda) = \frac{\lambda D}{2 + \lambda \left( D-2\right)} L_i (\btheta) \; ,
\end{eqnarray}
is the SLD in the noisy scenario. We obtain this expression by substituting $\rho_{\btheta}$ defined in~\eref{eq:statisticalModel1} in~\eref{eq:defLsolution}. From this result the QFI matrix $F(\btheta, \lambda)$ and the commutator matrix $A (\btheta, \lambda)$ are both found to be proportional to their noiseless counterparts $F(\btheta)$ and $A(\btheta)$ computed from $L_i(\btheta)$, i.e.
\begin{eqnarray}
	F(\btheta, \lambda) &=& \frac{\lambda^2 D}{2+\lambda \left( D-2 \right)} F (\btheta)\;, \\ A (\btheta, \lambda) &=& \frac{\lambda^3 D^2}{[2+ \lambda (D-1)]^2} A (\btheta) \;. 
\end{eqnarray}
Replaced into~(\ref{eq:Istar}) the above expressions lead to 
\begin{eqnarray}\label{rescaling} 
	\mathcal{I}^\star (\btheta, \lambda) = \frac{| \lambda| D}{2 +\lambda (D-2)} \mathcal{I}^{\star} (\btheta) \; ,
	\label{eq:IstarD}
\end{eqnarray}
with $\mathcal{I}^{\star} (\btheta)$ being the upper bound on the noiseless incompatibility figure of merit defined in~\eqref{eq:Istar}. Notice that this expression is not symmetric around $\lambda = 0$, i.e. $\mathcal{I}^\star (\btheta, \lambda) \neq \mathcal{I}^\star (\btheta, -\lambda)$ for $\lambda \ge 0$. We define
\begin{defn}(Asymmetry factor for $\mathcal{I}^\star (\btheta, \lambda)$)
	\begin{eqnarray}
	\kappa^\star (\lambda) := \frac{|\mathcal{I}^\star (\btheta, \lambda) - \mathcal{I}^\star (\btheta, -\lambda)| }{\mathcal{I}^\star (\btheta, \lambda)} = \frac{2 |\lambda| (D-2)}{2-\lambda (D-2)} \; .
	\end{eqnarray}
\end{defn}
The presence of an asymmetry in the properties of the $D$-dimensional depolarizing channel around $\lambda = 0$ was already pointed out in the context of communication in~\cite{Rosati2018}. For a qubit model $D = 2 \Longrightarrow \kappa^\star (\lambda) = 0$. We show through a numerical example that this asymmetry exists not only for the upper bound  $\mathcal{I}^\star (\btheta, \lambda)$ but also for the actual figure of merit  $\mathcal{I}(\btheta, \lambda)$. Consider the encoding of two near-zero phases ($d = 2$) on a qutrit ($D = 3$) via the unitary operator~\eref{eq:encodingDim} where the generators are chosen to be
\begin{eqnarray}
	H_1 = \begin{pmatrix}
	0 & - \rmi & 0 \\
	+ \rmi & 0 & 0 \\
	0 & 0 & 0
	\end{pmatrix}\;, \qquad  H_2 = \begin{pmatrix}
	1 & 0 & 0 \\
	0 & -1 & 0 \\
	0 & 0 & 0
	\end{pmatrix} \; ,
\end{eqnarray}
and the probe state is
\begin{eqnarray*}
	\ket{\psi} =  \frac{1}{\sqrt{2}} \begin{pmatrix} 1 \\ -1 \\ 0 \end{pmatrix} \; ,
\end{eqnarray*}
As in section~\ref{sec:threeQubit} the figure of merit has been computed with the semidefinite program presented in \ref{app:SDPprogram}, for $500$ equally spaced values of $\lambda$ in the allowed region. In figure~\ref{fig:asymmetry} the plot of $\mathcal{I}(\btheta = 0, \lambda)$ is reported for $\lambda \in \left( -1/8, 1 \right)$, with a zoom on $\lambda \in \left( -1/8, 1/8 \right)$. The dashed curve for $\lambda < 0$ reported in the insert is the reflection of the curve for $\lambda > 0$. It has been plotted in order to highlight the presence of the asymmetry. For this model, at the point $\btheta = 0$, the upper bound $\mathcal{I}^\star (\btheta = 0, \lambda)$ and the figure of merit $\mathcal{I} (\btheta = 0, \lambda)$ coincide. While in this qutrit example there is no gap between $\mathcal{I} (\btheta, \lambda)$ and $\mathcal{I}^\star  (\btheta, \lambda)$, in general for a $D$-dimensional model this could be the case. From~\eqref{eq:IstarD} we see that  
\begin{eqnarray}
	\lim_{\lambda \rightarrow 0} \mathcal{I}^\star \left(\btheta, \lambda \right) = 0 \; \Longrightarrow 	\lim_{\lambda \rightarrow 0} \mathcal{I} \left(\btheta, \lambda \right) = 0 \; ,
\end{eqnarray}
which means $\mathcal{I}^\star \left(\btheta, \lambda \right) -  \mathcal{I} \left(\btheta, \lambda \right) \rightarrow 0$ for $\lambda \rightarrow 0$.

\section{Design of compatible models for quantum metrology}
\label{sec:enforcing}
The following section is somewhat disconnected from the previous discussions on the incompatibility measure $\mathcal{I} (\btheta, \lambda)$. Here we want to analyze some strategies that have been proposed in the past and some generalizations that allow to produce a fully compatible statistical model in quantum metrology. We will only need to asses the condition $A(\btheta) = 0$ to claim compatibility, according to the bound~\eqref{eq:Istar}, and therefore we won't need the linear program for $\mathcal{I} (\btheta)$. We first review what is already known for $2$ qubits, each codified via~\eqref{eq:encoding}, and then generalize it for the $D$-dimensional encoding~\eqref{eq:encodingDim} when a couple of $D$-dimensional systems are available.

\subsection{Known results for a two-qubits probe}
\label{sec:twoQubits}
In this section we analyze the compatibility of three different two qubit encoding scenarios in the absence of noise and for some special instances of the input states. We do not claim paternity of these results, they are only reviewed here in order to be generalized in the next section. Consider first the ancilla-aided model, in which only one of the two qubits is subject to the unitary encoding \eqref{eq:encoding}. This means that the total evolution of the two qubits is $\id \otimes U_{{\btheta}}$. As input state for the probes we take a Bell state, which is known in the literature to be optimal for the estimation of $SU(2)$ operations~\cite{Fujiwara2001}, for example
\begin{eqnarray}
\ket{\psi} := \frac{\ket{00} + \ket{11}}{\sqrt{2}} \; ,
\label{eq:ancilla}
\end{eqnarray} 
where $\ket{0}$ and $\ket{1}$ are the eigenvectors of $\sigma_z$. From a direct computation we see that for this state $A({\btheta}) = 0 \; \forall \, \btheta$, which from~\eref{eq:Istar} gives
\begin{eqnarray}\label{LACCOMP} 
\mathcal{I}({\btheta}) =0 \;, \qquad \forall \, \btheta \;. 
\end{eqnarray} 
leading to compatibility. This result was first reported in~\cite{Fujiwara2001}. Interestingly enough compatibility can also be obtained when operating on the maximally entangled state~\eref{eq:ancilla} with the encoding $U_{{\btheta}} \otimes U_{{\btheta}}$. Indeed by explicit computation we get again $A({\btheta})=0 \; \forall \, \btheta$ that leads once more to (\ref{LACCOMP}). {Such result can be found in~\cite{Ragy2016}}. We will see in section~\ref{sec:ancillaInc} that these effects are just a special instance of a more general trend since a maximally entangled state of a finite dimensional probe always gives full compatibility, both for one and two uses of the encoding unitary channel. 

We now give a last example, which we here name ``anti-parallel spin strategy'' for future reference. Take the input state going through the encoding $U_{\btheta} \otimes U_{\btheta}$ to be $\ket{+ \widehat{\boldsymbol{n}}}\otimes \ket{- \widehat{\boldsymbol{n}}}$, where $\ket{+\widehat{\boldsymbol{n}}}$ and $\ket{-\widehat{\boldsymbol{n}}}$ have opposite Bloch vectors $+\widehat{\boldsymbol{n}}$ and $-\widehat{\boldsymbol{n}}$. This state has the same QFI of the state of two parallel spins $\ket{+\widehat{\boldsymbol{n}}}\otimes \ket{+\widehat{\boldsymbol{n}}}$, but has $A({\btheta}) = 0 \; \forall \btheta$ (in contrast to $\ket{+\widehat{\boldsymbol{n}}} \otimes \ket{+\widehat{\boldsymbol{n}}}$), which means that it is fully compatible and a superior probe for the sensing task. This result can be obtained from direct computation or thanks to the observation of section~\ref{sec:antiparalleD}, where we generalize this ideas to finite dimensional probes. The superiority of the anti-parallel spin state was already observed by Gisin and Popescu in~\cite{Gisin1999} and in the context of parameter estimation in~\cite{Lina2014}. In all these three examples, being the encoded state pure, a measure entangled across two qubits only is sufficient to get compatibility~\cite{Matsumoto2020}.

\subsection{Compatibility of the maximally entangled states}
\label{sec:ancillaInc}
In this subsection we will show that the results of section~\ref{sec:twoQubits} are only a particular case of a general observation, by proving that the use of an ancilla, maximally entangled with the probe, can completely remove the incompatibility, leading to the identity
\begin{eqnarray}\label{FULLCO} 
	\mathcal{I} (\btheta) = 0 \;, \qquad \forall \, \btheta\;,
\end{eqnarray}
Consider hence as input the following pure state
\begin{eqnarray}
	\ket{\psi} := \frac{1}{\sqrt{D}} \sum_{i=1}^d \ket{i} \otimes \ket{i} \in \mathcal{H} \otimes \mathcal{H} \; ,
	\label{eq:maximD}
\end{eqnarray}
on which the evolution $\id \otimes U_{{\btheta}}$ acts to produce the output state 
\begin{eqnarray}
	\ket{\psi_{{\btheta}}}:=\id \otimes U_{{\btheta}}\ket{\psi} \;, 
	\label{ENCO1} 
\end{eqnarray}
(i.e. the Choi–Jamio\l kowski state of the channel $U_{{\btheta}}$~\cite{Jam1972, Jam1975}). Following section~\ref{sec:ddim} and~\ref{app:effective}, the associated symmetric logarithmic derivatives~\eref{eq:defLpure} of $\rho_{\btheta} := \ket{\psi_{\btheta}} \! \bra{\psi_{\btheta}}$, can be expressed as
\begin{eqnarray}
	L_k (\btheta) = \frac{2 \rmi}{D} \sum_{ij} \ket{i} \! \bra{j} \otimes \left( H_{k}^{\text{eff}}(\btheta) \ket{i} \! \bra{j} - \ket{i} \! \bra{j} H_{k}^{\text{eff}}(\btheta) \right) \;,
\end{eqnarray}
which lead to the following expressions for the $F(\btheta)$ and $A(\btheta)$ matrices:
\begin{eqnarray}
	F_{lm} (\btheta)&=& 2 \Tr \left( \lbrace H_{l}^{\text{eff}}(\btheta), H_{m}^{\text{eff}}(\btheta) \rbrace \right)/D\;,
 	\\ A_{lm} (\btheta) &=& - 2 \rmi \Tr \left( \left[ H_{l}^{\text{eff}}(\btheta), H_{m}^{\text{eff}}(\btheta) \right] \right)/D = 0 \;,
\end{eqnarray}
where in the last identity we used the fact that $\left[ H_{l}^{\text{eff}}(\btheta), H_{m}^{\text{eff}}(\btheta) \right]$ is a traceless operator. Accordingly the upper bound~\eref{eq:Istar} imposes (\ref{FULLCO}), hence the thesis: the addition of a sufficiently large ancilla permits to remove entirely the quantum incompatibility for the LAC measurements.

A similar result holds true also when we let evolve the maximally entangled state~\eref{eq:maximD} through $U_{\btheta} \otimes U_{\btheta}$. In this case~(\ref{ENCO1}) gets replaced by
\begin{eqnarray} 
	\ket{\psi_{{\btheta}}}:=U_{\btheta} \otimes U_{\btheta}\ket{\psi} \;, 
	\label{ENCO2}
\end{eqnarray}
which leads to
\begin{eqnarray} 
	F_{lm} (\btheta) &=& 8 \Tr \left( \lbrace H_{l}^{\text{eff}}(\btheta) , H_{m}^{\text{eff}}(\btheta) \rbrace \right)/D \; , \\
	A_{lm} (\btheta) &=& - 4 \rmi \Tr \left( \left[ H_{l}^{\text{eff}}(\btheta) , H_{m}^{\text{eff}}(\btheta) \right] \right)/D = 0 \; ,
\end{eqnarray} 
which gives again the full compatibility condition~(\ref{FULLCO}).

\subsection{Generalized anti-parallel spin strategy}
\label{sec:antiparalleD}
Now we generalize the ``anti-parallel spin'' strategy of section~\ref{sec:twoQubits}. Suppose that we only have two parameters to estimate ($d = 2$) and we take for probe the separable input state $\ket{\psi_1} \otimes \ket{\psi_2}$ that evolves through the mapping induced by $U_{\btheta} \otimes U_{\btheta}$. The sufficient condition for compatibility  $A (\btheta)=0$ can be expanded as
\begin{eqnarray}
	\langle \psi_1 | [H^{\text{eff}}_1 (\btheta), H^{\text{eff}}_2 (\btheta)] | \psi_1 \rangle + \langle \psi_2 | [H^{\text{eff}}_1 (\btheta), H^{\text{eff}}_2 (\btheta)] | \psi_2 \rangle = 0 \; .
	\label{eq:compCondition}
\end{eqnarray}
The operator $[ H_{1}^{\text{eff}}(\btheta), H_{2}^{\text{eff}}(\btheta)]$ is skew-hermitian and therefore is diagonalizable and has purely imaginary eigenvalues $\pm \rmi a_j$, where $a_j > 0$, for $j= 1, 2, \dots, \floor{D/2}$, each associated with an eigenvector $\ket{\pm \rmi a_j}$. If the dimension $D$ is odd, then we have an extra unique zero eigenvalue. Let's denote with $V$ the unitary operator that performs such diagonalization, i.e. $V^{\dagger} \left[ H_{1}^{\text{eff}}(\btheta), H_{2}^{\text{eff}}(\btheta) \right] V = \text{diag} \left( \pm \rmi a_1, \pm \rmi a_2, \cdots \right)$. Let's define $\mathcal{S} \subseteq \lbrace 1, 2, \dots, \floor{D/2} \rbrace$, then a couple of states that guarantees compatibility is
\begin{eqnarray}
	\ket{\psi_1} &:=& \frac{1}{|\mathcal{S}|} \sum_{j \in \mathcal{S}} \rme^{i \varphi^1_j} V \ket{\rmi (-1)^{s_j} a_j} \; , 
	\label{eq:geneNpsi1} \\
	\ket{\psi_2} &:=& \frac{1}{|\mathcal{S}|} \sum_{j \in \mathcal{S}} \rme^{i \varphi^2_j} V \ket{- \rmi (-1)^{s_j} a_j} \; ,
	\label{eq:geneNpsi2}
\end{eqnarray}
where $\varphi_j^1$ and $\varphi_j^2$ are arbitrary phases, $s_j \in \lbrace 0, 1\rbrace$, and $|\mathcal{S}|$ is the cardinality of $\mathcal{S}$. Notice that we are also free to add in the definition of whichever $\ket{\psi_1}$ or $\ket{\psi_2}$ the state $V \ket{0}$, with $\ket{0}$ being the eigenvector with null eigenvalue (in case there is one). With the above choice of probes the compatibility condition~\eref{eq:compCondition} is realized. Notice also that the QFI matrix of $\ket{\psi_1} \otimes \ket{\psi_2}$ is the sum of the QFIs of $\ket{\psi_1}$ and $\ket{\psi_2}$. This two states taken individually manifest incompatibility, but when measured jointly they gain full compatibility. This construction is the analogue of the ``anti-parallel spin strategy'' of section~\eref{sec:twoQubits}. We observe that a state $\ket{\psi}$, being an equal superposition of $\ket{\psi_1}$ and $\ket{\psi_2}$ is also fully compatible. The condition~\eref{eq:compCondition} justifies also the compatibility of $\ket{+\widehat{\boldsymbol{n}}} \otimes \ket{-\widehat{\boldsymbol{n}}}$ $\forall \, \btheta$ in section~\ref{sec:twoQubits}. These two states are an orthogonal basis for the qubit Hilbert space, therefore 
\begin{eqnarray}
	\langle +\widehat{\boldsymbol{n}} | [H_1^{\text{eff}} (\btheta), H_2^{\text{eff}} (\btheta)] | + \widehat{\boldsymbol{n}} \rangle + \langle -\widehat{\boldsymbol{n}} | [H_1^{\text{eff}} (\btheta), H_2^{\text{eff}} (\btheta)] | -\widehat{\boldsymbol{n}} \rangle \nonumber \\ = \Tr \left( [H_1^{\text{eff}} (\btheta), H_2^{\text{eff}} (\btheta)] \right) = 0 \; , \nonumber 
\end{eqnarray}
hence condition~\eref{eq:compCondition} is satisfied and (\ref{FULLCO}) holds. Incidentally we observe also that for a $d$-parameter estimation, the state $\ket{\psi_1} \otimes \ket{\psi_2} \otimes \cdots \otimes \ket{\psi_D}$ is compatible when $\lbrace \ket{\psi_i} \rbrace_{i = 1}^D$ is base for the Hilbert space $\mathcal{H}$, because then we would have 
\begin{eqnarray}
	A_{ij} (\btheta) \propto \Tr \left( \left[ H_i^{\text{eff}} (\btheta), H_j^{\text{eff}} (\btheta) \right] \right) = 0 \; .
\end{eqnarray}

\section{Conclusions}
\label{sec:conclusions}
One of the defining properties of quantum mechanics is the intrinsic incompatibility between the possible experiments that could be carried out on a quantum system. This causes the appearance of information limits on the precision to which different characteristics of a certain quantum system can be known. Formally, this comes always from the non-commutativity of quantum operations. The main role of this paper is to give a theoretical foundation to the measure of incompatibility in the quantum estimation task. For this purpose we define in section~\ref{sec:figOfMerit} a figure of merit having a well defined operational and geometrical meaning. The figure of merit $\underline{r} (\btheta)$ in~\eref{eq:defr} is built to be easily liked to the asymptotic results of estimation theory~\cite{Hayashi2006}. This allows us to easily compute it, as explained in \ref{app:SDPprogram}, and to give the upper bound in section~\ref{sec:boundr}. The definition of incompatibility depends on the operations that we are able to perform, which is our level of control over the system. If we are only able to perform separable measurements on the probes then the relevant incompatibility measure is $\underline{r}^\text{s} (\btheta)$ defined in~\eqref{eq:defrs}. In section~\ref{sec:noisyInc} the estimation is studied in the scenario where a depolarizing noise acts, this is a form of disturbance which is often used to model the decoherence dynamics in metrology~\cite{Dob2012, Kolo2013}. We observed some interesting phenomena like the asymmetry of the incompatibility for inversion in the space of states in section~\ref{sec:ddim}. In section~\ref{sec:enforcing} we discuss some strategies able to produce compatible models for quantum metrology with generic $D$-dimension systems, which are for example the use of maximally entangled states. As a further development it would be interesting to determine which state gives the maximum and minimum incompatibility for a certain encoding at a fixed point $\btheta$. This optimization is hard because the figure of merit is a complicated non linear function of the state. In a sense the probe that maximizes incompatibility is the one which captures at most the quantum properties of the encoding process. Finally we would like to look for a link between the incompatibility and the Heisenberg scaling. In this context the only relevant measures are the one that assume no constraints on the separability of the input, because a single giant entangled probe would be used.

\section*{ACKNOWLEDGMENTS}
We thank Angelo Carollo,  Francesco Albarelli, Marco Genoni, and Yuxiang Yang for discussions. We acknowledges support by MIUR via PRIN 2017 (Progetto di Ricerca di Interesse Nazionale): project QUSHIP (2017SRNBRK).

\appendix

\section{Figure of merit for LAC measurements}
\label{app:LACfig}
In this appendix we will define a version of the incompatibility figure of merit for local asymptotic covariant (LAC) measurements~\cite{Hayashi2006, Yang2019}. Consider the sequences of POVMs $\left( {\text{M}}_k \right)_{k\in\mathbb{N}_0} \in \mathcal{M}_{\mathcal{C}}^{\text{\tiny{(LAC)}}}$ that satisfy local asymptotic covariance (LAC) at the point ${\btheta} \in \Theta$, as defined in~\cite{Hayashi2006, Yang2019}. The $k$th measurement ${\text{M}}_k$ of a sequence in $\mathcal{M}_{\mathcal{C}}^{\text{\tiny{(LAC)}}}$ acts on $k$ probes and has $\Sigma^{(k)}({\btheta})$ as associated MSE matrix. Because of the definition of LAC, $\left( {\text{M}}_k \right)_{k\in\mathbb{N}_0} \in \mathcal{M}_{\mathcal{C}}^{\text{\tiny{(LAC)}}}$ admits a limiting MSE matrix, i.e. $\lim_{k \rightarrow \infty} k \Sigma^{(k)}({\btheta}) := \Sigma({\btheta})$ (we also ask $\norm{\Sigma({\btheta})} < \infty$). The new figure of merit is hence defined as
\begin{eqnarray}
	r({\btheta}) := \inf_{\mathcal{M}_{\mathcal{C}}^{\text{\tiny{(LAC)}}}} \sup_{G \ge 0} \frac{\Tr \left[ G \cdot \Sigma({\btheta}) \right]}{\Tr \left[ G \cdot F^{-1}({\btheta})\right]} \; .
	\label{eq:alternativeDefinition}
\end{eqnarray}
In~\cite{Yang2019} the authors have proven the validity of the Holevo bound for the LAC measurements and its achievability in the same class (for a full rank $\rho_{{\btheta}}$ with non degenerate spectrum), that is
\begin{eqnarray}
	\inf_{\mathcal{M}_{\mathcal{C}}^{\text{\tiny{(LAC)}}}} \Tr \left[G \cdot \Sigma({\btheta}) \right] \; = C_{\text{H}} \left(G, {\btheta} \right) \; .
	\label{eq:holevoLAC}
\end{eqnarray}
We now prove that th $\inf$ and $\sup$ of~\eqref{eq:alternativeDefinition} can be commuted. It is easy to prove that the set $\mathcal{M}_{\mathcal{C}}^{\text{\tiny{(LAC)}}}$ is convex. The set of POVMs acting on ${\mathfrak{S}}\left( \mathcal{H}^{\otimes N} \right)$ and having ${\btheta}$ as outcome set can be thought as a convex subset of a certain dual vector space $\mathcal{V}'$~\cite{Chiribella20072}. The set containing the infinite sequences of $\mathcal{ V}'$ is a vector space, and the set of sequences thereof that are sequences of POVMs is convex. Furthermore the defining properties of LAC~\cite{Yang2019} is stable under convex combination of two measurement sequences. In the Minimax theorem of~\cite{Kneser1952} the both spaces are required to be be locally convex. Banach spaces like $\mathcal{G}$ and $\mathcal{V}' \supset \mathcal{M}_N$ are locally convex, and the countable infinite product space of multiple $\mathcal{V}'$ (which is the space of sequences) is also a locally convex space. Furthermore given $\Sigma^{1} (\btheta)$ and $\Sigma^{2} (\btheta)$ the limiting MSE matrices of two sequences $\left( {\text{M}}^1_k \right)_{k\in\mathbb{N}_0}$ and $\left( {\text{M}}^2_k \right)_{k\in\mathbb{N}_0}$, the asymptotic MSE of the convex combination $\left( \alpha {\text{M}}^1_k + \left( 1 - \alpha \right) {\text{M}}^2_k \right)_{k\in\mathbb{N}_0}$ is $ \Sigma (\btheta) = \alpha \Sigma^{1} (\btheta) + \left(1 - \alpha \right) \Sigma^{2} (\btheta)$. Consequently, just like in \ref{app:alternativeDefinition}, the Minimax theorem of Kneser~\cite{Kneser1952, Frenk2004} can be applied to swap the order of $\inf$ over ${\mathcal{M}_{\mathcal{C}}^{\text{\tiny{(LAC)}}}}$ and $\sup_{G \ge 0}$. It is understood that the argument of~\eref{eq:alternativeDefinition} can be cast into the same form of~\eref{eq:argument}, from which the linearity in $\left( {\text{M}}_k \right)_{k\in\mathbb{N}_0}$ and $G$, and the continuity in $G$ easily follow. We arrive therefore at
\begin{eqnarray}
	r({\btheta}) = \sup_{G \ge 0} \inf_{\mathcal{M}_{\mathcal{C}}^{\text{\tiny{(LAC)}}}} \frac{\Tr \left[ G \cdot \Sigma({\btheta}) \right]}{\Tr \left[ G \cdot F^{-1}({\btheta})\right]} = \sup_{G \ge 0} \frac{C_\text{H} (G,{\btheta})}{C_\text{S} (G,{\btheta})} = \underline{r} (\btheta)\;.
	\label{eq:finalExpressionr}
\end{eqnarray}

\section{Exchanging $\sup$ and $\inf$ in the figure of merit definition}
\label{app:alternativeDefinition}
We will arrive at~\eref{eq:CMI} through a series of lemmas.\\

\noindent \textit{Lemma 1}. The function $r_N \left( G, {\text{M}}_N , {\btheta} \right)$ is continuous in $G \in \mathcal{G}$ at fixed ${\text{M}}_N \in \mathcal{M}_N^{\text{(LU)}}$.\\

\noindent
The denominator of $r_N \left( G, {\text{M}}_N , {\btheta} \right)$ can be bounden as $\Tr \, [ G \cdot F(\btheta)^{-1} ] \ge \lambda_{\text{min}} \left( F(\btheta)^{-1} \right)= \lambda_{\text{max}} \left( F (\btheta) \right)^{-1} = 1/\norm{F(\btheta)}$, this means
\begin{eqnarray}
	r_N \left( G, {\text{M}}_N , {\btheta} \right) \le N \norm{F(\btheta)} \Tr \, [ G \cdot \cov (\btheta) ] \; , 
	\label{eq:normBound}
\end{eqnarray}
it follows
\begin{eqnarray*}
	| r_N \left( G_1, {\text{M}}_N, \btheta \right) - r_N \left( G_2, {\text{M}}_N, \btheta \right) | \le N \norm{F(\btheta)} \norm{\cov (\btheta)} \norm{G_1 - G_2} \; .
\end{eqnarray*}
Its important to assume that the set ${\text{M}}_N\in\mathcal{M}_N^{{\text{(LU)}}}$ contains only measurements with bounded MSE matrices, i.e. $\norm{{\cov} (\btheta)} \le C$. Therefore we have
\begin{eqnarray}
	| r_N \left( G_1, {\text{M}}_N, \btheta \right) - r_N \left( G_2, {\text{M}}_N, \btheta \right) | \le N C \norm{F (\btheta)} \norm{G_1 - G_2} \; ,
\end{eqnarray}
which means the function $r_N \left( G, {\text{M}}_N, \btheta \right)$ is Lipschitz continuous with constant $N C \norm{F(\btheta)}$ and therefore continuous. It will be useful in the latter to notice that $r_N ( F(\btheta)^\frac{1}{2} G F(\btheta)^\frac{1}{2}, {\text{M}}_N, \btheta) = N \Tr \, [ G \cdot F(\btheta)^{\frac{1}{2}} {\cov} (\btheta) F(\btheta)^{\frac{1}{2}} ]$ is also continuous in $G \in \mathcal{G}$, because is the composition of the continuous functions $r_N (G, {\text{M}}_N, \btheta)$ and $G \rightarrow F(\btheta)^\frac{1}{2} G F(\btheta)^\frac{1}{2}$.\\

\noindent 
In this paper only the upper semicontinuity of $r_N (G, {\text{M}}_N, \btheta)$ is actually used.\\

\noindent  \textit{Lemma 2}. The figure of merit $r_N (\btheta)$ defined in~\eref{eq:definitionR} can be equivalently expressed as
\begin{eqnarray}
	r_N (\btheta) = \sup_{G \ge 0} \inf_{{\text{M}}_N\in\mathcal{M}_N^{{\text{(LU)}}}}	r_N \left( G, {\text{M}}_N, \btheta \right) \; .
	\label{eq:definitionRalt}
\end{eqnarray}
\\
\noindent 
This Lemma is based on the application of the Minimax theorem of Kneser~\cite{Kneser1952, Frenk2004}. First of all we need to cast $r_N (\btheta)$ in a suitable form. We start from the observation that the set of positive weight matrices $G$ is invariant under congruence for the positive matrix $F(\btheta)^{\frac{1}{2}}$, i.e.
\begin{eqnarray}
	\lbrace G \ge 0 \rbrace = F(\btheta)^{\frac{1}{2}} \lbrace G \ge 0 \rbrace F(\btheta)^{\frac{1}{2}} = \lbrace F(\btheta)^{\frac{1}{2}} G F(\btheta)^{\frac{1}{2}} \ge 0 \rbrace \; .
\end{eqnarray}
This means that the $\sup$ can be taken on the matrices $F(\btheta)^{\frac{1}{2}} G F(\btheta)^{\frac{1}{2}} \ge 0$ without changing $r_N(\btheta)$, so we have
\begin{eqnarray}
	r_N (\btheta) &=& \inf_{{\text{M}}_N\in\mathcal{M}_N^{{\text{(LU)}}}} \sup_{F(\btheta)^{\frac{1}{2}} G F(\btheta)^{\frac{1}{2}} \ge 0} \frac{N \Tr \, [ F(\btheta)^{\frac{1}{2}} G F(\btheta)^{\frac{1}{2}} \cdot \cov (\btheta)]}{\Tr \, [ F(\btheta)^{\frac{1}{2}} G F(\btheta)^{\frac{1}{2}} \cdot F(\btheta)^{-1} ]} \\
	&=& \inf_{{\text{M}}_N\in\mathcal{M}_N^{{\text{(LU)}}}} \sup_{G \ge 0} \frac{N \Tr \, [ G \cdot F(\btheta)^{\frac{1}{2}} \cov (\btheta) F(\btheta)^{\frac{1}{2}} ]}{\Tr \, [ G ]} \nonumber \\
	&=& \inf_{{\text{M}}_N\in\mathcal{M}_N^{{\text{(LU)}}}} \sup_{G \in \mathcal{G}} N \Tr \, [ G \cdot F(\btheta)^{\frac{1}{2}} \cov (\btheta) F(\btheta)^{\frac{1}{2}} ] \; .
	\label{eq:argument}
\end{eqnarray}
In the second line, in the domain of the supremum, we have again used that every $F(\btheta)^{\frac{1}{2}} G F(\btheta)^{\frac{1}{2}} \ge 0$ corresponds to a $G \ge 0$. In the last equation the $\sup$ is restricted to the set $\mathcal{G} = \lbrace G \ge 0, \Tr \, G = 1 \rbrace$, which is compact and convex. The set of locally unbiased measurements ${\text{M}}_N\in\mathcal{M}_N^{{\text{(LU)}}}$ can be thought as a (non-empty) subset of a dual vector space~\cite{Chiribella20071, Chiribella20072}, which is a convex set because the locally unbiased measurements are stable under convex combination. The function $N \Tr \, [ G \cdot F(\btheta)^{\frac{1}{2}} \cov (\btheta) F(\btheta)^{\frac{1}{2}} ]$ is linear in both its arguments. The linearity in $G$ is self evident, so we only show linearity in the measurement. Suppose that we are given two POVMs denoted by ${\text{M}}^{1}$ and ${\text{M}}^2$, characterized respectively by the effects $\text{E}^1_{{\boldsymbol{\hat{\theta}}}}$ and $\text{E}^2_{{\boldsymbol{\hat{\theta}}}}$ associated to the outcome ${\boldsymbol{\hat{\theta}}}$. We have dropped for simplicity the subscript $N$ in ${\text{M}}^{1}$ and ${\text{M}}^2$ and we will also drop the superscript $(N)$ in the MSE matrix $\Sigma (\btheta)$. Consider the POVM being the linear combination ${\text{M}} := \lambda {\text{M}}^1 +\left( 1 - \lambda \right) {\text{M}}^2$. By definition its effects are
\begin{eqnarray}
	\text{E}_{{\boldsymbol{\hat{\theta}}}} := \lambda 	\text{E}^1_{{\boldsymbol{\hat{\theta}}}} + \left( 1 -\lambda \right) \text{E}^2_{{\boldsymbol{\hat{\theta}}}} \; .
\end{eqnarray}
The $\Sigma(\btheta)$ matrix associated to ${\text{M}}$ is computed as expectation value on the probability distribution 
\begin{eqnarray}
	p ({\boldsymbol{\hat{\theta}}}) &:=& \Tr ( \rho_{{\btheta}} \text{E}_{{\boldsymbol{\hat{\theta}}}} ) \nonumber \\ &=& \alpha \Tr (\rho_{{\btheta}} \text{E}^1_{{\boldsymbol{\hat{\theta}}}}) + ( 1 -\alpha ) \Tr ( \rho_{{\btheta}} \text{E}^2_{{\boldsymbol{\hat{\theta}}}}) \nonumber \\ &=& \alpha p^1 ({\boldsymbol{\hat{\theta}}}) + (1-\alpha) p^2 ({\boldsymbol{\hat{\theta}}}) \; .
\end{eqnarray}
The linearity of $p ( {\boldsymbol{\hat{\theta}}} )$ translates to the linearity of $\Sigma (\btheta)$, i.e.
\begin{eqnarray}
	\Sigma_{ij} (\btheta) &=& \alpha \Sigma^{1}_{ij} (\btheta) + (1 - \alpha) \Sigma^{2}_{ij} (\btheta)\; .
\end{eqnarray}
This means that the whole argument of the $\inf \sup$ in~\eref{eq:argument} is linear in the POVM, and it is additionally upper semicontinuous in $G$ at fixed ${\text{M}}_N$ (In \textit{Lemma 1} we proved continuity, which implies upper semicontinuity). We have now proved all the required hypothesis for the Minimax theorem of Kneser~\cite{Kneser1952, Frenk2004}, which allows us to write
\begin{eqnarray}
	r_N (\btheta) = \sup_{G \ge 0} \inf_{{\text{M}}_N\in\mathcal{M}_N^{{\text{(LU)}}}} \frac{N \Tr \left[ G \cdot \cov (\btheta) \right]}{\Tr \left[ G \cdot F(\btheta)^{-1}\right]} \; .
	\label{eq:definitionRminimax}
\end{eqnarray}
It is worth stressing that without such argument the quantity 
\begin{eqnarray}
	{r}_N^{(\downarrow)} (\btheta) := \sup_{G \ge 0} \inf_{{\text{M}}_N\in\mathcal{M}_N^{{\text{(LU)}}}} \frac{N \Tr \left[ G \cdot \cov (\btheta) \right]}{\Tr \left[ G \cdot F(\btheta)^{-1}\right]}
\end{eqnarray}
is by construction always smaller than or equal to 
\begin{eqnarray}
	r_N^{(\uparrow)} (\btheta) := \inf_{{\text{M}}_N\in\mathcal{M}_N^{{\text{(LU)}}}} \sup_{G \ge 0} \frac{N \Tr \left[ G \cdot \cov (\btheta) \right]}{\Tr \left[ G \cdot F(\btheta)^{-1}\right]} \; ,
\end{eqnarray}	
i.e. ${r}_N^{(\downarrow)} (\btheta) \leq r_N ^{(\uparrow)} (\btheta)$.\\

\noindent \textit{Lemma 3}.
\begin{eqnarray}
	r_{N_1} (\btheta) \ge r_{N_1 N_2} (\btheta) \; .
\end{eqnarray}
\\
\noindent
Let $N := N_1 N_2$ with $N_1$, $N_2$ integers. Given $N$ copies of the probe we can organize them into $N_2$ distinct subgroups, each of them containing $N_1$ probes. We now perform exactly the same measurement ${\text{M}}_{N_1}$ on each group and use the $N_2$ outcomes to estimate ${\btheta}$ by taking their arithmetic mean. Calling this measurement ${\text{M}}^\star_{N}$ it follows that its MSE matrix $\Sigma^\star_N (\btheta)$ corresponds to $\Sigma_{N_1} (\btheta) /N_2$, being $\Sigma_{N_1} (\btheta)$ the MSE matrix of ${\text{M}}_{N_1}$. This holds true because the estimators are unbiased at ${\btheta}$. Therefore we have 
\begin{eqnarray}
	r_{N}(G, {\text{M}}^\star_{N}, \btheta) &=& \frac{N \Tr \left[ G \cdot \Sigma^\star_N (\btheta) \right]}{\Tr \left[ G \cdot F(\btheta)^{-1}\right]} \\ &=& \frac{N \Tr \left[ G \cdot \Sigma_{N_1} (\btheta) \right]}{N_2 \Tr \left[ G \cdot F(\btheta)^{-1}\right]} \\ &=& \frac{N_1 \Tr \left[ G \cdot \Sigma_{N_1} (\btheta) \right]}{\Tr \left[ G \cdot F(\btheta)^{-1}\right]} \\ &=& r_{N_1}(G, {\text{M}}_{N_1}, \btheta) \; .
\end{eqnarray}
We now need to introduce a new quantity:
\begin{eqnarray}
	r_N (G, \btheta) := \inf_{{\text{M}}_N\in\mathcal{M}_N^{{\text{(LU)}}}} r_N (G, {\text{M}}_N, \btheta) \; , 
	\label{eq:defRNG}
\end{eqnarray}
where the supremum on $G \ge 0$ has been removed. We can always take ${\text{M}}_{N_1}^{\varepsilon}$ such that 
\begin{eqnarray}
	r_{N_1} (G, {\text{M}}_{N_1}^{\varepsilon}, \btheta) \le \inf_{{\text{M}}_{N_1}\in\mathcal{M}_{N_1}^{{\text{(LU)}}}} r_{N_1} \left(G, {\text{M}}_1, \btheta \right) + \varepsilon \; ,
\end{eqnarray}
with $\varepsilon > 0$. Then we have
\begin{eqnarray}
	r_N \left(G, \btheta \right) &\le& r_{N}(G, {\text{M}}^\star_{N}, \btheta) = r_{N_1} (G, {\text{M}}_{N_1}^{\varepsilon}, \btheta) \nonumber \\
	&\le& \inf_{{\text{M}}_{N_1}\in\mathcal{M}_{N_1}^{{\text{(LU)}}}} r_{N_1} \left( G, {\text{M}}_{N_1}, \btheta \right) + \varepsilon \nonumber \\
	&=& r_{N_1} \left(G, \btheta \right) + \varepsilon \; . \nonumber 
\end{eqnarray}
where $r_N (G, \btheta)$ has been defined in~\eref{eq:defRNG}. Because of the arbitrariness of $\varepsilon$ we have $r_{N} \left(G, \btheta \right) \le r_{N_1} \left( G, \btheta \right)$. Taking $\sup_{G \ge 0}$ on both the members of this inequality gives finally the thesis.\\

\noindent \textit{Lemma 4.}
\begin{eqnarray}
	\underline{r} (\btheta) = \inf_{N \ge 1} r_N (\btheta) = \lim_{n \rightarrow \infty} \min_{N \le n} r_N (\btheta)\; .
	\label{eq:lemma2}
\end{eqnarray}
\\
\noindent 
Recall the definition of $\underline{r} (\btheta)$ in~\eref{eq:defr}, it can be expressed as $\underline{r} (\btheta) = \lim_{n\rightarrow \infty} \underline{r}_n (\btheta)$, where
\begin{eqnarray} \label{de} 
	\underline{r}_n (\btheta) :=\inf_{N \geq n} r_N (\btheta)\;, 
\end{eqnarray}
which is by construction non-decreasing in $n$, i.e.
\begin{eqnarray}
	\underline{r}_{m} (\btheta) \geq \underline{r}_n (\btheta) \qquad \qquad \forall \, m\geq n \;. 
\end{eqnarray}
Our goal is to show that due to \textit{Lemma 3}, the inequality in the above expression is always saturated, or equivalently that
\begin{eqnarray}
	\underline{r}_{m} (\btheta) = \underline{r}_1 (\btheta) = \inf_{N \geq 1} r_N (\btheta) \qquad \forall m \, \geq 1 \;, \label{ddfs}
\end{eqnarray}
which will lead automatically to $\underline{r} (\btheta) = \underline{r}_1 (\btheta) = \inf_{N \ge 1} r_N (\btheta)$. We can prove~(\ref{ddfs}) by contradiction: assume that there exists $m$ such that $ \underline{r}_{m} (\btheta) > \underline{r}_1 (\btheta)$. This implies that there must exist $k< m$ integer such that
\begin{eqnarray}
	\underline{r}_{1} (\btheta) = r_k (\btheta) < \underline{r}_m (\btheta) \;.
\end{eqnarray} 
This however can't be true because thanks to \textit{Lemma 3} we must have
\begin{eqnarray} 
	r_k (\btheta) \geq r_{mk} (\btheta) \geq \inf_{N \geq m} r_N (\btheta) = \underline{r}_m (\btheta) \;. 
\end{eqnarray}
For the second equality in~\eref{eq:lemma2} it is easy to notice that the sequence defined by 
\begin{eqnarray}
	\underline{r}^{(<)}_n (\btheta) := \min_{N \le n} r_N (\btheta) \; ,
	\label{eq:definitionr<}
\end{eqnarray}
is explicitly decreasing, i.e.
\begin{eqnarray}
	\underline{r}_{n+1}^{(<)} (\btheta) \le \underline{r}_{n}^{(<)} (\btheta) \;,
\end{eqnarray}
therefore its limit exists and it is easy to show it coincides with $\inf_{N \ge 1} r_N (\btheta)$ as we see in the following. Take $\varepsilon > 0$, then $\exists \, N_{\varepsilon}$ such that
\begin{eqnarray}
	r_{N_{\varepsilon}} (\btheta) \le \inf_{N \ge 1} r_N (\btheta) + \varepsilon \; ,
\end{eqnarray}
therefore $\forall \, \varepsilon$, $\exists \, n_{\varepsilon} := N_{\varepsilon}$ such that 
\begin{eqnarray}
	\underline{r}^{(<)}_{n_\varepsilon} (\btheta) \le r_{N_{\varepsilon}} (\btheta) \le \inf_{N \ge 1} r_N (\btheta) + \varepsilon \; ,
	\label{eq:upp}
\end{eqnarray}
furthermore
\begin{eqnarray}
	\underline{r}_n^{(<)} (\btheta) \ge \inf_{N \ge 1} r_N (\btheta) \; ,
	\label{eq:down}
\end{eqnarray}
because in $\underline{r}_n^{(<)} (\btheta)$ the domain of minimization is smaller. Together~\eref{eq:upp} and~\eref{eq:down} mean 
\begin{eqnarray}
	\lim_{n \rightarrow \infty} \underline{r}_n^{(<)} (\btheta) = \inf_{N \ge 1} r_N (\btheta) = \underline{r}_1 (\btheta) \; .
\end{eqnarray}
Before proceeding further let us make yet another observation: the above construction can be applied even in the absence of the optimization over $G$. Specifically, for all $G \ge 0$ and $N \in \mathbb{N}_0$, we define the the quantities
\begin{eqnarray}
	\underline{r}_n (G, \btheta) &:=& \inf_{N \ge n} r_N (G, \btheta) \; , \label{eq:defRnG} \\
	\underline{r}_n^{(<)} (G, \btheta) &:=& \min_{N \le n} r_N (G, \btheta) \; ,
\end{eqnarray}
with $r_N (G, \btheta)$ defined in~\eref{eq:defRNG}. By construction, for all given $G$, $\underline{r}_n (G, \btheta)$ is explicitly non-decreasing, while $\underline{r}_n^{(<)} (G, \btheta)$ is explicitly non-increasing, i.e.
\begin{eqnarray}
	\underline{r}_{n+1} (G, \btheta) &\ge& \underline{r}_n (G, \btheta) \; , \\
	\underline{r}_{n+1}^{(<)} (G, \btheta) &\le& \underline{r}_{n}^{(<)} (G, \btheta) \; .
	\label{eq:r<decreasing}
\end{eqnarray}
Moreover following the same arguments presented in \textit{Lemma 4} it turns out that $\underline{r}_n (G, \btheta)$ is indeed constant, i.e.
\begin{eqnarray}
	\underline{r}_n (G, \btheta) = \underline{r}_1 (G, \btheta) \; \quad \quad \forall n \; .
\end{eqnarray}
and
\begin{eqnarray}
	\underline{r}_1 (G, \btheta) &=& \inf_{N \ge 1} r_N (G, \btheta) \\ &=& \lim_{n \rightarrow \infty} \min_{N \le n} r_N (G, \btheta) \\ &=& \lim_{n \rightarrow \infty} \underline{r}_{n}^{(<)} (G, \btheta) \; .
	\label{eq:simplification}
\end{eqnarray}
\\

\noindent  \textit{Lemma 5}. The function $r_N \left(G, \btheta \right)$ is upper semicontinuous in $G \in \mathcal{G}$ $\forall \, N$.\\

\noindent 
The function $r_N \left( G, {\text{M}}_N, \btheta \right)$, defined in~\eref{eq:preTrivial}, is continuous in $G \in \mathcal{G}$ for fixed ${\text{M}}_N$ because of \textit{Lemma 1}, and in particular upper semicontinuous. The function $r_N (G, \btheta)$ defined in~\eref{eq:defRNG} is upper semicontinuous in $G \in \mathcal{G}$ because the infimum of a family of upper semicontinuous functions (here labeled by the measurement ${\text{M}}_N$) is upper semicontinuous.\\ 

\noindent
Consider next the supremum over $G$ of $\underline{r}_1 (G, \btheta)$, this can be evaluated as
\begin{eqnarray}
	\sup_{G \ge 0} \underline{r}_1 (G, \btheta) &=& \sup_{G \ge 0} \inf_{N \ge 1} r_N (G, \btheta) \label{eq:firstLine} \\ &=& \sup_{G \ge 0} \inf_{N \ge 1} \inf_{{\text{M}}_N\in\mathcal{M}_N^{{\text{(LU)}}}} r_N (G, {\text{M}}_N, \btheta) \nonumber \\ &=& \inf_{N \ge 1} \inf_{{\text{M}}_N\in\mathcal{M}_N^{{\text{(LU)}}}} \sup_{G \ge 0} r_N (G, {\text{M}}_N, \btheta) \nonumber \\ &=& \inf_{N \ge 1} r_N (\btheta) = \underline{r}_1 (\btheta) = \underline{r} (\btheta) \label{eq:secondLine} \; .
	\label{eq:evaluatedSup}
\end{eqnarray}
Going from~\eref{eq:firstLine} to~\eref{eq:secondLine} requires the application of two different versions of the Minimax theorem. First of all we need to commute $\inf_{N \ge 1}$ and $\sup_{G \ge 0}$ ($\sup_{\mathcal{G}}$). This can be accomplished with the Ky Fan Minimax theorem~\cite{KyFan1953, Frenk2004}. In order to use this result it must be proved that $r_N (G, \btheta)= \inf_{{\text{M}}_N\in\mathcal{M}_N^{{\text{(LU)}}}} r_N (G, {\text{M}}_N, \btheta)$ is Ky Fan concave-convex on $\mathcal{G} \times \mathbb{N}_0$. This condition is equivalent to having Ky Fan concavity in $G$ for every fixed $N$ and Ky Fan convexity in $N$ for every fixed $G$. Let us fix an arbitrary $N \in \mathbb{N}_0$. Given the combination $G_\alpha := \alpha G_1 + \left( 1 - \alpha \right) G_2$ with $\alpha \in \left[0, 1 \right]$, $\forall \, \varepsilon$ $\exists \, {\text{M}}_N^{\varepsilon}$ such that
\begin{eqnarray}
	r_N \left(G_\alpha, \btheta \right) &=& \inf_{{\text{M}}_N\in\mathcal{M}_N^{{\text{(LU)}}}} r_N \left( G_\alpha, {\text{M}}_N, \btheta \right) \\ &\ge& r_N \left( G_\alpha, {\text{M}}_N^{\varepsilon}, \btheta \right) - \varepsilon \; .
	\label{eq:concr1}
\end{eqnarray}
By expanding $G_{\alpha}$ in $r_N \left( G_\alpha, {\text{M}}_N^{\varepsilon}, \btheta \right) $ we have
\begin{eqnarray}
	 r_N \left( G_\alpha, {\text{M}}_N^{\varepsilon}, \btheta \right) = \alpha \, r_N \left( G_1, {\text{M}}_N^{\varepsilon}, \btheta \right) + \left(1-\alpha \right) r_N \left( G_2, {\text{M}}_N^{\varepsilon}, \btheta \right) \; ,
\end{eqnarray}
which thanks to the definition of $\inf_{{\text{M}}_N\in\mathcal{M}_N^{{\text{(LU)}}}}$ becomes
\begin{eqnarray*}
	 \fl \quad \quad r_N \left( G_\alpha, {\text{M}}_N^{\varepsilon}, \btheta \right) \ge \alpha \inf_{{\text{M}}_N\in\mathcal{M}_N^{{\text{(LU)}}}} r_N(G_1, {\text{M}}_N, \btheta) + \left( 1 - \alpha \right) \inf_{{\text{M}}_N\in\mathcal{M}_N^{\text{(LU)}}} r_N(G_2, {\text{M}}_N, \btheta) \; ,
\end{eqnarray*}
finally, substituting $r_N (G, \btheta)$, we get
\begin{eqnarray}
	r_N \left( G_\alpha, {\text{M}}_N^{\varepsilon}, \btheta \right) \ge \alpha r_{N} \left(G_1, \btheta \right) + \left(1 -\alpha \right) r_{N} \left( G_2, \btheta \right) \; .
	\label{eq:concr2}
\end{eqnarray}
Putting together~\eref{eq:concr1} and~\eref{eq:concr2} gives
\begin{eqnarray}
	r_{N} \left( G_\alpha, \btheta \right) \ge \alpha r_N \left( G_1, \btheta\right) + \left( 1 - \alpha \right) r_{N} \left(G_2, \btheta\right) - \varepsilon \; .
\end{eqnarray}
which for $\epsilon \rightarrow 0$ is the (Ky Fan) concavity condition for $r_N \left(G, \btheta \right)$. Let's now prove the Ky Fan convexity in $N$. Consider $N_1, N_2 \in \mathbb{N}_0$ and an arbitrary $G \in \mathcal{G}$, we have
\begin{eqnarray}
	 r_{N_1 N_2} \left(G, \btheta \right) \le \alpha \, r_{N_1} \left(G, \btheta \right) + (1-\alpha) r_{N_2} \left( G, \btheta \right) \, \forall \, \alpha \in \left[ 0, 1 \right] \; .
\end{eqnarray}
This is true because thanks to \textit{Lemma 3} we have $r_{N_1 N_2} (G, \btheta) \le r_{N_1} (G, \btheta)$ and $r_{N_1 N_2} (G, \btheta) \le r_{N_2} (G, \btheta)$. \textit{Lemma 5} proves that $r_N \left(G, \btheta \right)$ is upper semicontinuous in $G$ for every fixed $N$, this concludes the hypothesis check for the application of the Ky Fan Minimax theorem, according to which
\begin{eqnarray}
	\sup_{\mathcal{G}} \inf_{N \ge 1} r_{N} \left(G, \btheta \right) = \inf_{N \ge 1} \sup_{\mathcal{G}} r_{N} \left(G, \btheta \right) \; .
\end{eqnarray}
In order to get~\eref{eq:secondLine} from~\eref{eq:firstLine} we still need
\begin{eqnarray}
	\sup_{\mathcal{G}} \inf_{{\text{M}}_N\in\mathcal{M}_N^{\text{(LU)}}} r_N \left( G, {\text{M}}_N, \btheta \right) = \inf_{{\text{M}}_N\in\mathcal{M}_N^{\text{(LU)}}} \sup_{\mathcal{G}} r_N \left( G, {\text{M}}_N, \btheta \right) \; .
\end{eqnarray}
This is the content of \textit{Lemma 2}. By putting together~\eref{eq:simplification} and~\eref{eq:evaluatedSup} we get the expression
\begin{eqnarray}
	\underline{r} (\btheta) = \sup_{G \ge 0} \lim_{n \rightarrow \infty} \underline{r}_n^{(<)} (G, \btheta) \; .
\end{eqnarray}
Expanding this expression we arrive at~\eqref{eq:CMI}, with $C(G, \btheta)$ defined in~\eqref{DEFCFTHETA}.
%s

\section{Formulation of the semidefinite program}
\label{app:SDPprogram}
We start from~\eqref{eq:hcrbound} and write
\begin{eqnarray}
	\underline{r} (\btheta) &=& \sup_{G \ge 0} \frac{C_{\text{H}} \left(G, \btheta \right)}{\Tr \left[G \cdot F(\btheta)^{-1} \right]} \nonumber \\ &=& \sup_{G \in \mathcal{G}} C_{\text{H}} \left( F(\btheta)^{\frac{1}{2}} G F(\btheta)^{\frac{1}{2}}, \btheta \right)\, .
	\label{eq:startNum}
\end{eqnarray}
The semidefinite program for $C_{\text{H}} \left(G, \btheta \right)$ reported in~\cite{Albarelli2019} is
\begin{eqnarray}
	\begin{aligned}
	C_{\text{H}} \left(G, \btheta \right)\; = \; & \underset{V \in \mathbb{S}^n, X \in \mathbb{R}^{\tilde{d} \times n}}{\text{minimize}}
	& & \Tr \, [G \cdot V] \\
	& \text{subject to}
	& & \begin{pmatrix}
	V & X^T R_{\btheta}^\dagger\\
	R_{\btheta} X & \id_{\tilde{r}}
	\end{pmatrix} \ge 0 \; ,\\
	&&& X^T \frac{\partial \boldsymbol{s}_{\btheta}}{\partial {\btheta}} = \id_n \; .
	\end{aligned}
\end{eqnarray}
See the work~\cite{Albarelli2019} for the definitions of all the objects appearing in this program, it is not necessary to understand them in order to follow our derivation. Equation~\eref{eq:startNum} becomes
\begin{eqnarray}
	\begin{aligned}
	\underline{r} (\btheta) \; = \; & \sup_{G \in \mathcal{G}} \underset{V \in \mathbb{S}^n, X \in \mathbb{R}^{\tilde{d} \times n}}{\text{minimize}}
	& & \Tr \, [F(\btheta)^{\frac{1}{2}} G F(\btheta)^{\frac{1}{2}} \cdot V] \\
	& \text{subject to}
	& & \begin{pmatrix}
	V & X^T R_{\btheta}^\dagger\\
	R_{\btheta} X & \id_{\tilde{r}}
	\end{pmatrix} \ge 0 \; ,\\
	&&& X^T \frac{\partial \boldsymbol{s}_{\btheta}}{\partial \btheta} = \id_n \; .
	\end{aligned}
\end{eqnarray}
The objective function $\Tr \, [F(\btheta)^{\frac{1}{2}} G F(\btheta)^{\frac{1}{2}} \cdot V]$ is linear and continuous in both $G$ and $V$. The domain of the $\sup$ and the $\min$ are both convex, with $\mathcal{G}$ being compact. We can therefore apply again the Mimimax theorem of~\cite{Kneser1952} as done in \ref{app:alternativeDefinition}. Having $\sup_{G \in \mathcal{G}}$ as the innermost operation we can solve it and write
\begin{eqnarray}
	\sup_{G \in \mathcal{G}} \Tr \, [ F(\btheta)^{\frac{1}{2}} G F(\btheta)^{\frac{1}{2}} \cdot V ] = \norm{F(\btheta)^{\frac{1}{2}} V F(\btheta)^{\frac{1}{2}}} \; .
\end{eqnarray}
Now the objective of the minimization is the spectral norm of $F^{\frac{1}{2}}(\btheta) V F^{\frac{1}{2}}(\btheta)$. We can introduce a dummy variable $t$ and write the program as
\begin{eqnarray}
	\begin{aligned}
	\underline{r} (\btheta) \; = \; & \underset{V \in \mathbb{S}^n, X \in \mathbb{R}^{\tilde{d} \times n}}{\text{minimize}}
	& & t \\
	& \text{subject to}
	& & \norm{F (\btheta)^{\frac{1}{2}} V F(\btheta)^{\frac{1}{2}}} \le t \; , \\
	& & &\begin{pmatrix}
	V & X^T R_{\btheta}^\dagger\\
	R_{\btheta} X & \id_{\tilde{r}}
	\end{pmatrix} \ge 0 \; ,\\
	&&& X^T \frac{\partial \boldsymbol{s}_{\btheta}}{\partial \btheta} = \id_n \; .
	\end{aligned}
\end{eqnarray}
The condition on $\norm{F(\btheta)^{\frac{1}{2}} V F(\btheta)^{\frac{1}{2}}}$ can be written as
\begin{eqnarray*}
	\begin{aligned}
		& \lambda_{\max} ( F(\btheta)^{\frac{1}{2}} V F (\btheta) V F(\btheta)^{\frac{1}{2}} ) \le t^2 \Longrightarrow \\
		& F(\btheta)^{\frac{1}{2}} V F(\btheta) V F(\btheta)^{\frac{1}{2}} \le t^2 \id \Longrightarrow \\
		& t \id - F(\btheta)^{\frac{1}{2}} V F(\btheta)^{\frac{1}{2}} \left( t \id \right)^{-1} F(\btheta)^{\frac{1}{2}} V F(\btheta)^{\frac{1}{2}} \ge 0 \; .
	\end{aligned}
\end{eqnarray*}
Because of the Schur complement condition for the positive semidefinite matrices~\cite{Zhang2005} the optimization becomes
\begin{eqnarray}
	\begin{aligned}
	\underline{r} (\btheta) \; = \; & \underset{\substack{V \in \mathbb{S}^n, \\ X \in \mathbb{R}^{\tilde{d} \times n}}}{\text{minimize}}
	& & t \\
	& \text{subject to}
	& & \begin{pmatrix}
	t \id & F(\btheta)^{\frac{1}{2}} V F(\btheta)^{\frac{1}{2}}\\
	F(\btheta)^{\frac{1}{2}} V F(\btheta)^{\frac{1}{2}} & t \id
	\end{pmatrix} \ge 0 \; , \\
	& & & \begin{pmatrix}
	V & X^T R_{\btheta}^\dagger\\
	R_{\btheta} X & \id_{\tilde{r}}
	\end{pmatrix} \ge 0 \; ,\\
	&&& X^T \frac{\partial \boldsymbol{s}_{\btheta}}{\partial \btheta} = \id_n \; .
	\end{aligned}
	\label{eq:finalSDP}
\end{eqnarray}
From which we compute $\mathcal{I} (\btheta)$ according to~\eref{eq:defI}. This semidefinite program is solved by means of the modeling system CVX developed on Matlab~\cite{CVX}. 

\section{Explicit computation of $\underline{r}^\star (\btheta)$}
\label{app:proofrstar}
In this section we prove that $\sup_{G \ge0}$ in the definition~\eref{eq:usefulBound2} of $\underline{r}^\star (\btheta)$ can be computed exactly and we obtain the explicit expression for $\underline{r}^\star (\btheta)$ in~\eref{eq:explicitRstar}. First of all we define $A'(\btheta) := F(\btheta)^{-\frac{1}{2}} A(\btheta) F(\btheta)^{-\frac{1}{2}}$. This means we can write~\eref{eq:usefulBound2} as
\begin{eqnarray}
	\underline{r}^{\star} (\btheta) - 1 &=& \sup_{G \in \mathcal{G}} \Tr \abs \left[ G \cdot A' (\btheta) \right] \nonumber \\ &=& \sup_{G \in \mathcal{G}} \Tr | \sqrt{G} A' (\btheta) \sqrt{G}| \nonumber \\ &=& \sup_{G \in \mathcal{G}} \Tr \left[ \sqrt{\sqrt{G} \left( - A'(\btheta) G A'(\btheta) \right) \sqrt{G} }\right] \; ,
	\label{eq:quasiFidelity}
\end{eqnarray}
with the $\sup$ taken on $\mathcal{G}$. Because $A(\btheta)^\dagger = - A(\btheta)$ it holds that $-A(\btheta)'GA(\btheta)' \ge 0$. The trace in~\eref{eq:quasiFidelity} can be associated to the definition of the (squared) fidelity between the states identified by the matrices $G$ and $-A(\btheta)'GA(\btheta)'$. Notice that this last state must be normalized. Therefore we write
\begin{eqnarray}
	\underline{r}^{\star} (\btheta) -1 &=& \sup_{G \in \mathcal{G}} \sqrt{\mathcal{F}} \left( G , - A(\btheta)' G A(\btheta)' \right) \\ &=& \sqrt{\Tr \left[ -A(\btheta)' G A(\btheta)' \right]} \cdot \\ &\cdot& \sup_{G \in \mathcal{G}} \sqrt{\mathcal{F}} \left( G , \frac{- A(\btheta)' G A(\btheta)'}{\Tr \left[ - A(\btheta)' G A(\btheta)' \right]} \right) \nonumber \; .
\end{eqnarray}
We will prove that there is a choice of $G$ that gives both the maximum of $\Tr \left[ -A(\btheta)' G A(\btheta)' \right]$ and of the squared fidelity. Let's write $A(\btheta)'$ in the form $A(\btheta)' = Q M Q^T$ where $M$ is a block diagonal matrix having $2 \times 2$ blocks 
\begin{eqnarray*}
	M_i := \begin{pmatrix} 0 & \lambda_i \\ -\lambda_i & 0 \end{pmatrix} \; ,
\end{eqnarray*}
with $0 \le \lambda_{i+1} \le \lambda_{i} \in \mathbb{R}$. If $A'(\btheta)$ is of odd size the matrix $M$ has the last row and column full of $0$. We have $\Tr \left[ - A(\btheta)' G A(\btheta)' \right] = \Tr \, [- M \widetilde{G} M ]$, with $\widetilde{G} = Q^T G Q$, which explicitly reads
\begin{eqnarray*}
	\fl \quad \quad \quad \Tr \, [ - M \widetilde{G} M ] = \lambda_1^2 \left( \widetilde{G}_{11} + \widetilde{G}_{22} \right) + \lambda_2^2 \left( \widetilde{G}_{33} + \widetilde{G}_{44} \right) + \lambda_3^2 \left( \widetilde{G}_{55} + \widetilde{G}_{66} \right) + \cdots \; .
\end{eqnarray*}
The maximum of the above expression is $\lambda_1^2$, realized for a $\widetilde{G}$ having $\widetilde{G}_{11} + \widetilde{G}_{22} = 1$ and all the other matrix elements null. Notice that $\norm{F^{-\frac{1}{2}} A (\btheta) F^{-\frac{1}{2}}} = \norm{A(\btheta)'} = \lambda_1$. For the square fidelity to reach its maximum ($\sqrt{\mathcal{F}}= 1$) it must be
\begin{eqnarray}
	G = -\frac{A(\btheta)' G A(\btheta)'}{\Tr \left[ -A(\btheta)' G A(\btheta)' \right]} \;,
\end{eqnarray}
this is realized for $\widetilde{G}_{11} = \widetilde{G}_{22} = \frac{1}{2}$. Therefore we have build implicitly a matrix $G$ that saturates the $\sup$ and gives~\eref{eq:explicitRstar}.

\section{Effective generators for $\btheta \neq 0$}
\label{app:effective}
Consider the single qubit encoding given in~\eref{eq:encoding}, in order to compute the relevant metrological quantities, for example the QFI and the Holevo-Cram\'er-Rao bound, it is necessary to take the derivatives of this evolution, evaluated at $\btheta$, i.e. $\frac{\partial U_{\btheta}}{\partial \btheta} \Big |_{\btheta}$. If we are sensing small deviations of the phases around $\btheta = 0$, then these expressions are fairly easily computable, they are indeed $\frac{\partial U_{\btheta}}{\partial \theta_1} \Big |_{\btheta} = \rmi \sigma_y$, and $\frac{\partial U_{\btheta}}{\partial \theta_2} \Big |_{\btheta}= \rmi \sigma_z$. But if the base point of the sensing process is $\btheta \neq 0$, then these derivatives became cumbersome, and can hinder the derivation of simple analytical results. To overcome this issue we show in this section that the metrological properties of the estimation at a point $\btheta \neq 0$ are equivalent to that of a sensing process around zero, where the encoding has the effective generators $H_1^{\text{eff}} (\btheta)$ and $H_2^{\text{eff}} (\btheta)$, which are null-trace hermitian operators depending on the non null point $\btheta$ and are in general not simply $\sigma_y$ and $\sigma_z$. We write explicitly the small variations $\delta \btheta$ from the base point $\btheta$ in the encoding~\eref{eq:encoding}, i.e.
\begin{eqnarray}
	U_{{\btheta} + \delta {\btheta}} &:=& \exp \left( \rmi [(\theta_1 + \delta \theta_1) \sigma_y + (\theta_2 + \delta \theta_2) \sigma_z] \right) \\ &:=& \exp \left( \rmi H + \rmi \delta H \right) \; .
	\label{eq:altEncoding}
\end{eqnarray}
The variables $ \delta {\btheta}$ are now the unknown parameters, while $\btheta$ is known and fixed. The Hamiltonians $H := \theta_1 \sigma_y + \theta_2 \sigma_z$ and $\delta H := \delta \theta_1 \sigma_y + \delta \theta_2 \sigma_z$ have been defined. We expand the expression for $U_{\btheta + \delta \btheta}$ in terms of $H$ and $\delta H$ with the Baker-Campbell-Hausdorff formula, and keep only the first order terms in the infinitesimal variation $\delta \btheta$, obtaining
\begin{eqnarray}
	U_{{\btheta} + \delta {\btheta}} \simeq U_{{\btheta}} \exp \left( \rmi \delta H - \frac{1}{2} [\rmi H, \rmi \delta H] + \frac{1}{6} [\rmi H, [\rmi H, \rmi \delta H]] + \dots \right) \; .
\end{eqnarray}
Now, the idea is to perform the rotation $U_{-\btheta}$ on the probe after the encoding with $U_{\btheta + \delta \btheta}$, in such way we compensate for the know component of the rotation $U_{\btheta + \delta \btheta}$ and leave only a term depending on the new unknown variables $\delta \btheta$.
\begin{eqnarray}
	U_{-{\btheta}} U_{{\btheta} + \delta {\btheta}} &\simeq& \exp \left( \rmi \delta H - \frac{1}{2} [\rmi H, \rmi \delta H] + \frac{1}{6} [\rmi H, [\rmi H, \rmi \delta H]] + \dots \right) \label{eq:commutatorSeries}\\ &=& \exp \left( \rmi \left[ \delta \theta_1 H^{\text{eff}}_1 ({\btheta}) + \delta \theta_2 H^{\text{eff}}_2 ({\btheta}) \right] \right) \; .
	\label{eq:collection}
\end{eqnarray}
In the last expression we have collected the terms multiplied by $\theta_1$ and $\theta_2$ respectively, which have been named $H^{\text{eff}}_1 (\btheta)$ and $H^{\text{eff}}_2 (\btheta)$. Notice that the commutator of two skew-hermitian operators like $\rmi H$ and $\rmi \delta H$ is again skew-hermitian, this applies to all the elements of the exponentiated sum in~\eref{eq:commutatorSeries}, and means that the right hand side of~\eref{eq:commutatorSeries} is a unitary operator even if we have neglected higher order terms in $\delta \btheta$. The exponentiated sum is either equal to $\rmi \delta \theta_1 H^{\text{eff}}_1 ({\btheta})$ or to $\rmi \delta \theta_2 H^{\text{eff}}_2 ({\btheta})$ when we set either $\delta \theta_2 = 0$ or $\delta \theta_1 = 0$, therefore the effective generators are also hermitian operators. Consider a probe $\rho$ codified by $U_{\btheta + \delta \btheta}$, i.e. $\rho_{\btheta + \delta \btheta} \simeq U_{\btheta + \delta \btheta} \rho U_{\btheta + \delta \btheta}^\dagger$. All the informational quantities remain the same if a know unitary is applied to the state, indeed its effects can be always absorbed at the measurement stage (if the selected measurements set allows to do so). By choosing $U_{-\btheta}$ to be this unitary we get $U_{-\btheta} U_{\btheta + \delta \btheta} \rho U_{\btheta + \delta \btheta}^\dagger U_{-\btheta}^\dagger = U_{\delta \btheta} \rho U_{\delta \btheta}^\dagger$, with $U_{\delta \btheta} := U_{-\btheta} U_{\btheta + \delta \btheta}$. We observe that the traces of $H_1^{\text{eff}} (\btheta)$ and $H_1^{\text{eff}} (\btheta)$ can be neglected without consequences, indeed they contribute only to a global phase. Also if the gate $U_{\btheta}$ is used multiple times on an entangled state, so that the encoding is $U_{\btheta} \otimes U_{\btheta} \otimes \dots \otimes U_{\btheta}$, the traces of the generators only give an irrelevant global phase. We now further manipulate the encoding and look for a parameterization in which the generators are orthogonal. Two qubits operators $H_1$ and $H_2$ are said to be orthogonal if $\lbrace H_1, H_2 \rbrace = 0$. As null-trace hermitian operators on a qubit $H^{\text{eff}}_{1} (\btheta)$ and $H^{ \text{eff}}_{2} (\btheta)$ can be written
\begin{eqnarray}
	H^{\text{eff}}_{1} (\btheta) &=& \alpha_1 ({\btheta}) \sigma_x + \beta_1 ({\btheta}) \sigma_y + \gamma_1 ({\btheta}) \sigma_z \; , \label{eq:gen1} \\
	H^{\text{eff}}_{2} (\btheta) &=& \alpha_2 ({\btheta}) \sigma_x + \beta_2 ({\btheta}) \sigma_y + \gamma_2 ({\btheta}) \sigma_z \; , \label{eq:gen2}
\end{eqnarray}
with $\balpha(\btheta) := (\alpha_1 (\btheta), \alpha_2 (\btheta), \alpha_3 (\btheta)) \in \mathbb{R}^3$ and $\bbeta (\btheta) := (\beta_1 (\btheta), \beta_2 (\btheta), \beta_3 (\btheta)) \in \mathbb{R}^3$. The orthogonality condition is then $\lbrace H^{\text{eff}}_{1} (\btheta), H^{\text{eff}}_{2} (\btheta) \rbrace = 2 \balpha (\btheta) \cdot \bbeta (\btheta) \, \id$. We can decompose $H^{\text{eff}}_{2} (\btheta)$ in a term proportional to $H^{\text{eff}}_{1} (\btheta)$ and one orthogonal as following
\begin{eqnarray}
	H^{\text{eff}}_{2} (\btheta) := \frac{\balpha (\btheta) \cdot \bbeta (\btheta)}{\norm{\balpha (\btheta)}^2} H^{\text{eff}}_{1} (\btheta) + H^{\bot \text{eff}}_{2} (\btheta)\; ,
	\label{eq:defHorth}
\end{eqnarray}
this is the definition of $H^{\bot \text{eff}}_{2} (\btheta)$, which satisfies $\lbrace H^{\text{eff}}_{1} (\btheta), H^{\bot \text{eff}}_{2} (\btheta) \rbrace = 0$. We define $x(\btheta) := \balpha (\btheta) \cdot \bbeta (\btheta) / \norm{\balpha (\btheta)}^2$ for ease of notation and substitute~\eref{eq:defHorth} in~\eref{eq:collection}, thus getting
\begin{eqnarray}
	U_{\delta \btheta} = \exp \left( \rmi [(\delta \theta_1 + x({\btheta}) \delta \theta_2) H^{\bot \text{eff}}_{1} ({\btheta}) + \delta \theta_2 H^{\bot \text{eff}}_{2} ({\btheta})] \right) \; ,
\end{eqnarray}
where $H^{\text{eff}}_{1} (\btheta)$ has been renamed $H^{\bot \text{eff}}_{1} (\btheta)$. The final step is to normalize the generators, thus defining $\widetilde{H}^{\bot \text{eff}}_{1} ({\btheta}) := H^{\bot \text{eff}}_{1} ({\btheta}) / \Tr \left[ H^{\bot \text{eff}}_{1} ({\btheta})^2 \right]$, and $\widetilde{H}^{\bot \text{eff}}_{1} ({\btheta}) := H^{\bot \text{eff}}_{2} ({\btheta}) / \Tr \left[ H^{\bot \text{eff}}_{2} ({\btheta})^2 \right]$. Going from $H_i^{\text{eff}} (\btheta)$ to $\widetilde{H}^{\bot \text{eff}}_{i} ({\btheta})$ corresponds to the following reparametrization
\begin{eqnarray}
	\begin{cases}
		\delta \theta_1' = \Tr \left[ H^{\bot \text{eff}}_{2} ({\btheta})^2 \right] (\delta \theta_1 + x(\btheta) \delta \theta_2) \; , \\
		\delta \theta_2' = \Tr \left[ H^{\bot \text{eff}}_{2} ({\btheta})^2 \right] \delta \theta_2 \; .
	\end{cases}
\end{eqnarray}
A rotation of the reference system can align $\widetilde{H}^{\bot \text{eff}}_{i} ({\btheta})$ with $\sigma_y$ and $\sigma_z$, remember thought that the probe state must also be transformed. Let us introduce the unitary $V_{\btheta}$ such that $V_{\btheta} \widetilde{H}^{\bot \text{eff}}_{1} ({\btheta}) V_{\btheta}^\dagger = \sigma_y$ and $V_{\btheta} \widetilde{H}^{\bot \text{eff}}_{2} ({\btheta}) V_{\btheta}^\dagger = \sigma_z$, then $V_{\btheta} U_{\delta \btheta} V_{\btheta}^\dagger = \rme^{\rmi \left( \delta \theta_1' \sigma_y + \delta \theta_2' \sigma_z \right)}$, while the probe state becomes $V_{\btheta} \rho V_{\btheta}^\dagger$.


\begin{thebibliography}{100}
\bibitem{NOTA} Suppose in fact that the POVM ${\text M}_N$ produces outputs $\omega$ belonging to a numerical set $\Omega\neq \Theta$ which are then processed in order to build the estimator of ${{\btheta}}$ through a function ${\boldsymbol{\tilde{\theta}}}: \Omega \rightarrow \Theta$. The associated statistics can then be equivalently described by means of a new POVM $\{ E_{{\boldsymbol{\hat{\theta}}}}^{(N)}\}_{\boldsymbol{\hat{\theta}}}$ with output ${\boldsymbol{\hat{\theta}}} \in \Theta$ by setting
%
\begin{eqnarray}
	E_{{\boldsymbol{\hat{\theta}}}}^{(N)} = \int d\omega \delta({\boldsymbol{\hat{\theta}}} - {\boldsymbol{\tilde{\theta}}}(\omega)) Y_{\omega}\;, \nonumber 
\end{eqnarray}
%
where $\{ Y_{\omega} \}_{\omega\in \Omega}$ are the elements of the original POVM and where $\delta(\cdots)$ is the Dirac delta function.
\bibitem{NOTADEP} Notice that by construction the infimum over ${\text{M}}_N$ in~(\ref{eq:definitionR}) could in principle force the selected POVMs to bare a non trivial dependence upon the unknown value~${{\btheta}}$. This is a very well known fact in quantum metrology which is at the origin of a circular reasoning issue which in practice can be cured by adopting adaptive estimation approaches. In our case, however, the possible functional dependence of ${\text{M}}_N$ on ${{\btheta}}$ is not of particular concern as it simply reflects the local character of the figure of merit.
\bibitem{NOTAsingular} The parameterization $(\alpha, \beta)$ is singular at the points $\alpha = \frac{\pi}{2}$ and $\alpha = \frac{3 \pi}{3}$, and so are the matrices $F(\alpha, \beta)$ and $A(\alpha, \beta)$. In case $\boldsymbol{a}_{\btheta}$ exactly corresponds to one of these points we simply choose another set of spherical coordinates which have the singularity located at a different position.
\end{thebibliography}
\end{document}